\documentclass{svjour3}                     
\smartqed  
\usepackage{graphicx}
\begin{document}

\title{Fresnel Zone Plate Telescopes for X-ray Imaging II: Numerical simulations
with parallel and diverging beams
\thanks{This work was made possible in part from a grant from Indian Space Research Organization (ISRO).
SP and DD thank CSIR/NET scholarships which supported their research work.
}
}

\titlerunning{Simulations with Zone Plates}   

\author{S. Palit \and Sandip K. Chakrabarti \and D. Debnath \and A. R. Rao \and A. Nandi \and Vipin K. Yadav \and V. Girish}

\authorrunning{Chakrabarti et al. } 

\institute{S. Palit, D. Debnath, A. Nandi$^+$, V. Yadav$^+$ \at
Indian Centre for Space Physics, 43 Chalantika, Garia Station Rd., Kolkata 700084\\
Tel.: +91-33-24366003\\
              Fax: +91-33-24622153 Ext. 28\\
              \email{sourav@csp.res.in; dipak@csp.res.in; anuj@csp.res.in; vipin@csp.res.in}   
($+$: Posted at ICSP by Space Science Division, ISRO Head Quarters)
           \and
Sandip K. Chakrabarti \at
              S.N. Bose National Centre for Basic Sciences, JD Block, Salt Lake, Kolkata 700097 \\
(Also at Indian Centre for Space Physics, 43 Chalantika, Garia Station Rd., Kolkata 700084)\\
              Tel.: +91-33-23355706\\
              Fax: +91-33-23353477\\
              \email{chakraba@bose.res.in}           
\and
V. Girish, A.R. Rao \at
Tata Institute of Fundamental Research, Homi Bhabha Road, Colaba, 400025\\
              \email{giri@isac.gov.in; arrao@tifr.res.in}   
}
\date{Received: date / Accepted: date}
\maketitle

\begin{abstract}
We present the results of simulations of shadows cast by a zone plate telescope which 
may have one to four pairs of zone plates. From the shadows we reconstruct the 
images under various circumstances. We discuss physical basis of the resolution
of the telescope and demonstrate this by our simulations. We allow the source to be at a finite distance (diverging beam) as well as at an infinite distance (parallel beam)
and show that the resolution is worsened when the source is nearby. By reconstructing the 
zone plates in a way that both the zone plates subtend the same solid angles at 
the source, we obtain back high resolution even for sources at a finite distance. We
present simulated results for the observation of the 
galactic center and show that the sources of varying intensities may be reconstructed with accuracy.
Results of these simulations would be of immense use in interpreting 
the X-ray images from recently launched CORONAS-PHOTON satellite.

\end{abstract}

\section{Introduction}
\label{intro}

Zone Plate Telescopes (ZPTs) have generated immense theoretical interest in the past.
(Ables, 1968; Dicke, 1968; Barrett \& Swindell, 1996; Desai, Norris \& Nemiroff, 1993;
Desai et al. 1993; 1998, 2000). Due to the availability of technology to make finer zones
using X-ray and gamma-ray opaque materials, the interest of using zone plates have increased
recently. In Chakrabarti et al. (2009, hereafter Paper I), we presented extensive theoretical 
studies and some of the results of the experiments conducted at the Indian Centre for Space Physics 
X-ray laboratory on zone plate telescopes. Such telescopes have been used in Indian payload 
system RT-2 aboard the Russian satellite CORONAS-PHOTON which was launched on 30th January, 2009.
As mentioned in Paper I the ZPTs have an advantage over other high resolution 
X-ray telescopes in that they can have arbitrarily high angular resolution and that the
resolution can also be independent of energy bands in a large range of energy. The only disadvantage 
is of course that the ZPT  is a two element system as opposed to the
conventional coded aperture masks (CAMs) which are single element systems. 

In the present paper of our series we shall present the results of the Monte-Carlo simulations of the 
nature of photon distribution on the detector planes for various combinations of 
zone plates and X-ray sources. The basic starting point in this paper would be 
Paper I and references therein. Here, we consider one pair to four pairs of zone 
plate combinations, with a point source and distributed sources. 
We also vary the distance of source from finite to infinite. We hope to be able 
to present the most comprehensive results in this subject so that
interpretation of the data from CORONAS-PHOTON may become easier. In our next 
paper (Nandi et al. 2009) the instrumentation and early results from RT-2 payload which uses
ZPTs would be presented. 

The plan of the present paper is the following: In the next Section, 
we discuss the angular resolution of a ZPT and also the variation of the 
 nature of the reconstructed image when the source distance varies.
In Section 3, we will present in detail the results of our simulations in
various circumstances. Finally, in Section 4, we make concluding remarks.

\section{Theoretical considerations}

In Paper-I, we discussed the theory of the ZPTs and how the images are reconstructed.
We also discussed the situation when a pseudo-source or a DC-offset appears. Presently, 
we discuss the resolution and source broadening when the source distance is 
varied and how to rectify the broadening effect.

\subsection{Resolution of dual Fresnel zone plate telescopes for point sources}

Resolution of an instrument refers to its ability to distinguish two closely placed sources.
It is the minimum angular separation between two sources which could be separated. 
For a pair of aligned zone plates, this can be computed as follows (see Fig. 1):

\begin{figure}[h]
\includegraphics[height=1.50in,width=5.0in]{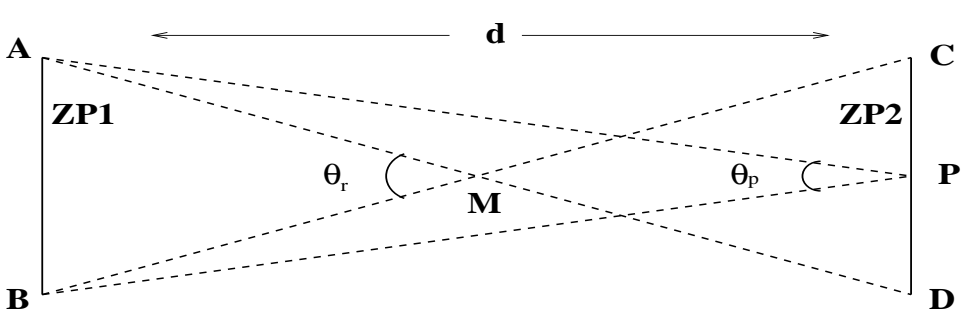}
\caption{Computation of the angular resolution of dual Fresnel zone plates. Photons from two successive 
transparent zones at A and B of the first plate ZP1 arrive at point P on an opaque zone CD of 
the second zone plate ZP2. The angular resolution is the angle $\theta_r$ since any point
in CD is equivalent for imaging purpose.}

\label{}
\end{figure}

Let AB and CD denote the finest zones in the first and the second zone plates (ZP1 and ZP2) respectively. 
A point $P$ on CD can distinguish two sources as `separate' sources only if photons from them arrive 
at $P$ through two different zones, say through points A and B, 
where AB is the opaque zone of the first zone plate ZP1. However, since any 
point in CD would be equivalent, the location of $P$ could vary from $C$ 
to $D$ and as such, the resolution would be governed not by the angle $\theta_p$, but by 
$$
\theta_r = 2\omega /d~~{\rm radian},
\eqno{(1)}
$$ 
where, $w  =(\sqrt{n}-\sqrt{n-1}) r_{in}$, width of the considered zone, $r_{in}$ being the inner radius 
and $n$ is the zone number under consideration. When we consider inner zones, the situation is worsened, 
as the width of the zone becomes larger.

In resolving the sources, and indeed, even to be able to identify a single source, the size of 
the detector pixel $d_p$ also plays a major role. For instance, if $d_p>w$ for all $n$, in the
computation of the resolution, $w$ has to be replaced by $d_p$. However, if the source 
is so off-axis that two successive Moir\'{e} fringes fall on the same detector pixel, 
then that source would not be `visible' by the telescope. To show this, 
let us recall from Paper I that the fringe separation $s$ in a 
zone plate is given by $s= r_{in}^2/r_{s}$, where, $r_{s}$ is the distance of the projection 
of the center of ZP1 on ZP2 as measured from the center of ZP2 and is given by $=d~\times~\theta$, 
$d$ being the separation between zone plates, and $\theta$ being the offset angle. The largest possible offset 
angle is obtained by equating $s/2$ with the pixel size $p$ of the detector. This gives 
$\theta_{max} = r_{in}^2/2pd$. For our set up in RT-2/CZT payload (Nandi et al., 2009),
$r_{in}=0.122$cm, $p=0.25$cm for CZT detector and $d=30$cm. This gives $\theta_{max}=3.41$ arcmin. Thus the 
field of view of the ZPT with a CZT detector would be at the most $\sim 7 $arc min. If the source
casts fringes along the diagonal of the zone plate, $p$ would be replaced by  $p/\sqrt{2}$ cm 
and the value of $\theta_{max}$ would be $4.82$ arc min. 
 
\subsection {Broadening of reconstructed point sources kept at finite distances}

So far, we discussed the resolution when the object is placed at a large distance. When the source is 
close by, the image is distorted since various points of a single transparent zone is located at various 
incident angles due to divergence of the beam. In Fig. 2 we show that photons from a source $S$ are
incident to form a Moir\'{e} fringe at various angles (say, $\theta_1$ and $\theta_2$) depending on the
location of the fringe. When the inverse transformation is made to reconstruct the source, it 
would assume the direction of the source as the directions of the incident rays in different points 
of the fringe. Thus the source is not exactly reproduced and generally the source 
is broadened. The reconstructed source is contained in the same solid angle as that 
produced by the Moir\'{e} pattern on the actual source (i.e., that obtained by the largest angular size 
$\theta_1+\theta_2$). When the source distance is increased, this solid angle becomes zero 
and a point source is reproduced as a point (if permissible by the detector pixel). 
In case the Moir\'{e} pattern is not circular, but has some other shape (e.g., like a gibbous moon, 
see Fig. 9 below), the sum $\theta_1+\theta_2$ will be direction dependent 
and hence the reconstructed image of a point source
would also be like a miniature version of the Moir\'{e} pattern itself.

\begin{figure}[h]
\includegraphics[height=1.50in,width=5.0in]{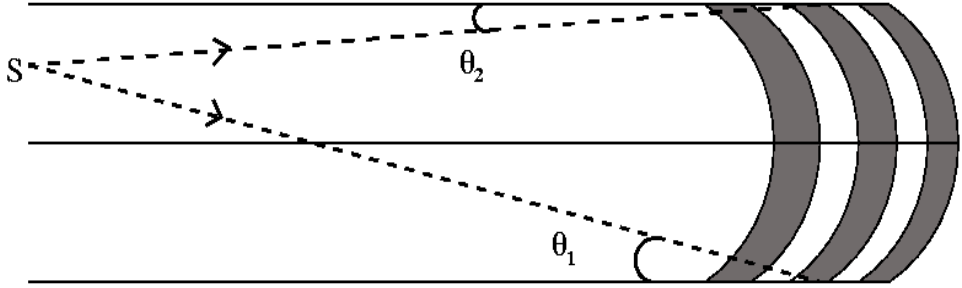}
\caption{The cause of broadening and deformation  of a point source at a finite distance
while reconstruction is explained. Incident photons subtend unequal angles on the Moir\'{e}
pattern. The finite solid angle, which could be direction dependent because of specific
Moir\'{e} pattern causes broadening of the image.}
\label{}
\end{figure}

\subsection {Resolution of the sources kept at finite distances}

The broadening of the reconstructed image clearly affects the resolution by worsening it.
To see this, let us assume that the width of the outermost transparent zone is $w$ as before, and the 
zone plates are separated by a distance $d$. The angular resolution is $\theta_r$ (Eq. 1). This 
becomes meaningful only if the reconstructed source has an angular size less than $\theta_r$. Let $Z$
be the average distance of the sources at an angular distance of $\theta_r$ (Fig. 1) 
and the outer radius of the front zone plate be $r_{out}$, then from Fig. 2, the angular size 
of the Moir\'{e} pattern as seen by an average source point $\theta_R$  (= $\theta_1+\theta_2$)
is given by $2r_{out}/Z$. In order that broadened images are also resolved, we must have 
$ \theta_r \geq  \theta_R$ i.e.,
$$
Z \geq \frac{d r_{out}}{\omega}= \frac{d \sqrt{n}}{\sqrt{n}-\sqrt{n-1}} .
\eqno{(3)}
$$
It is to be noted that the result is independent of the size of the zone plate since the 
result only depends on the number of zones and neither on $r_{in}$ nor on $r_{out}$. 
As an example, for zone plates with number of zones, $n=150$ and separated by $d=20$~cm, the distance $Z$
must be at least $\sim 6000$~cm in order that the resolution remains at least as good as the point source.
Closer sources will have a resolution inferior to this.

\subsection{Improvement of the resolution for a source at a finite distance}

A simpler way to improve the resolution when the source is at a finite distance would be 
to reconstruct the telescope in such a way that both the zone plates subtend equal solid angle 
 as seen from the source. This can be easily achieved by changing the radii of the zones of the
second zone plate in such a way that the radius of the original inner ring $r_{in, orig}$ is replaced by
$$
r_{in}=r_{in,orig} (1+d/Z)
$$
where, $d$ is the distance between the zone plates and $Z$ is the distance of ZP2 from the source. One can 
recompute the resolution of such a ZPT and easily show that the resolution would be the same as the 
original zone plate telescope having the source at infinity.

\section{Results of Monte-Carlo simulations}

\subsection{Effects of number of zone plates in the Telescope} 

So far, we presented some basic theoretical results. In what follows, we present results of simulations
of diverse kinds in order to show that the theoretical considerations hold true.
We will generally consider zone plates having $151$ zones with the innermost radius $r_{in}=0.122$~cm.
However, the distance $d$ may vary. This will be mentioned as and when necessary.
In Paper~I, it was discussed that one requires four pairs of zone plates in a complete
X-ray imaging telescope in order that neither the DC-offset, nor the pseudo-source (alias)
appears in the reconstructed image. Furthermore, one also has very less noise in the 
reconstructed image, in general. In this subsection, we will discuss the results of simulations 
and discuss under what condition this holds true.

\subsubsection{Four pairs of zone plates}

In this Section, we consider all the four pairs placed in the following way:

\noindent 1st quadrant elements:      $\sqrt(n)$  \&     $\sqrt(n)$  (positive and negative respectively)\\
\noindent 2nd quadrant elements:      $\sqrt(n)$ \&  $\sqrt(n-0.5)$ (positive and negative respectively)\\
\noindent 3rd quadrant elements:      $\sqrt(n)$ \&  $\sqrt(n-0.5)$ (both positive)\\
\noindent 4th quadrant elements:      $\sqrt(n)$ \&   $\sqrt(n)$ (both positive)

Here, by $\sqrt{n}$ type, we mean the plate having $n$th zone radius $r_{n}=\sqrt{n} r_{in}$. 

First, we carry out the simulations with the source at infinity. The zone plate 
spacing in each case is taken to be $10$ cm. The source is placed at an angular 
distance of $\phi=1500$ arcsec from the optical axis and at a zenith angle
of $\theta=45$ deg measured from the positive X-axis. The number of photons infalling 
on each of the front zone plates (ZP1s) is $10^5$. The fringes obtained in each of the pairs 
is given in Fig. 3a. The source obtained by reconstruction is shown in Fig. 3b. 
Note that neither the DC offset, nor the pseudo-source appears in the reconstructed image.

\begin{figure}[h]

\includegraphics[height=2.5in,width=2.5in]{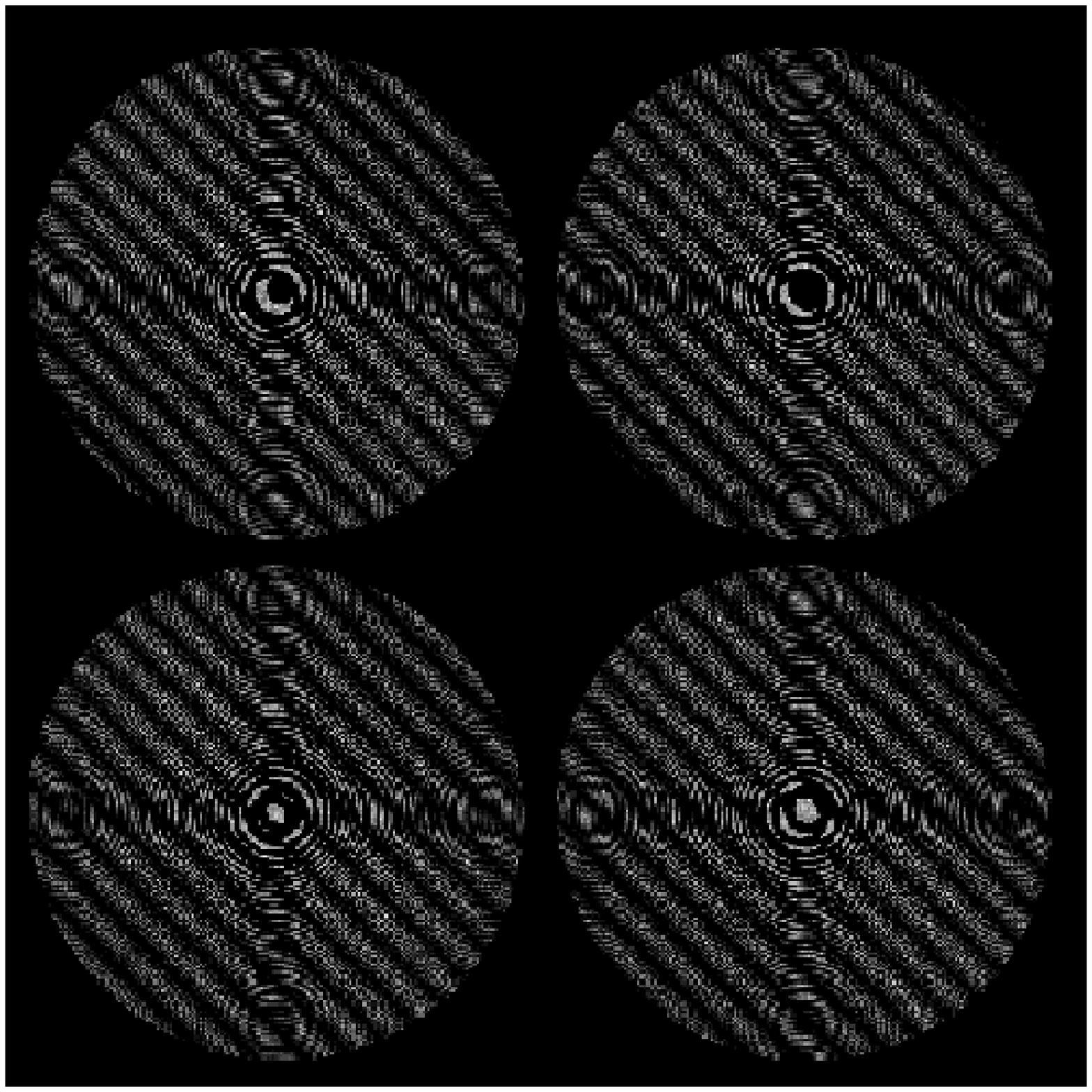}\hspace{0.0cm}
\includegraphics[height=2.3in,width=2.3in]{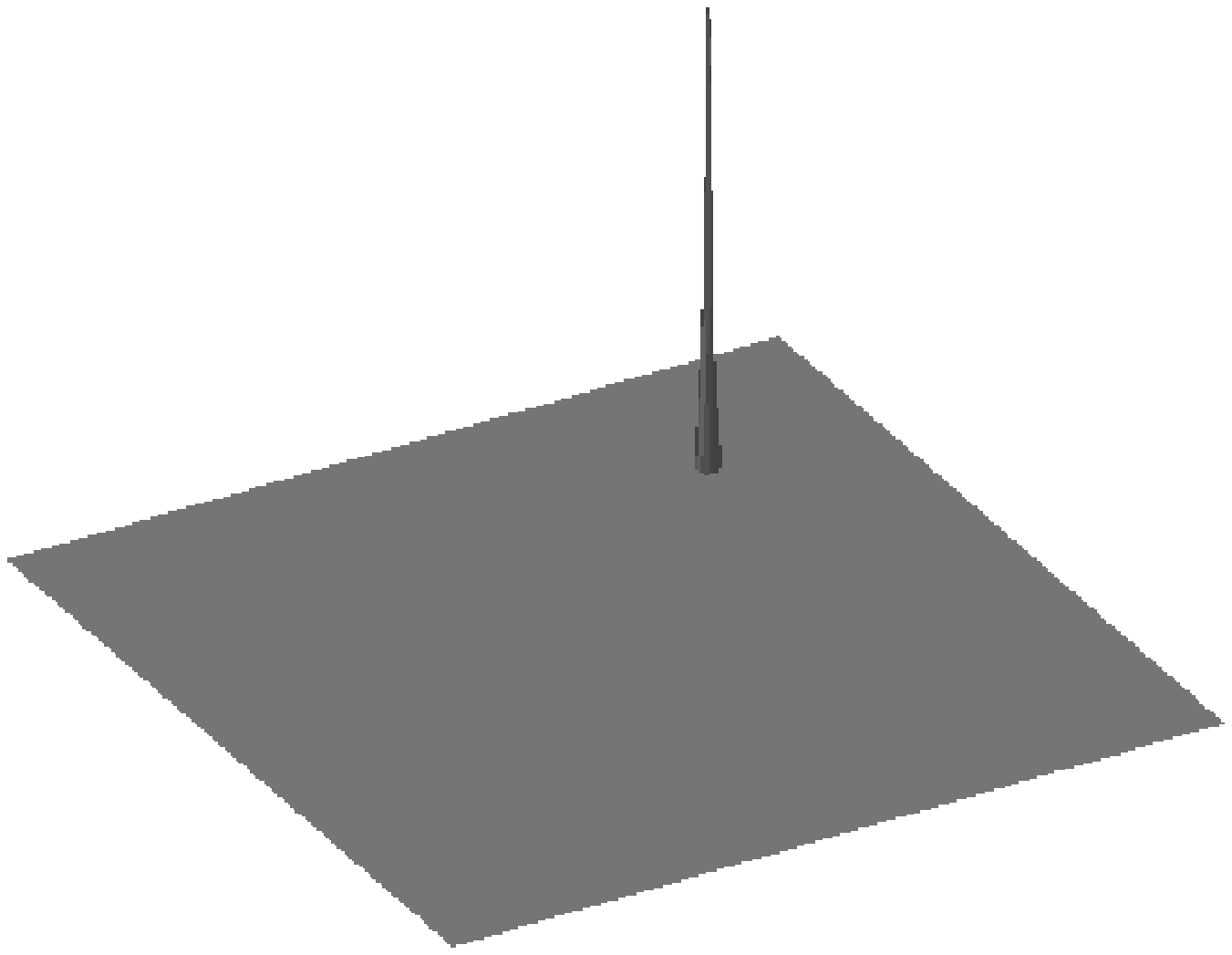}
\caption{(a) Fringes obtained for a source at an infinite distance with four pairs of zone plates.
(b) Reconstructed source on a CMOS detector ($4500$ arcsec along each side of the squared base. 
Neither the pseudo-source nor the DC offset are seen in the reconstructed image.}
\end{figure}

If the source is brought to a finite distance, however, the DC off-set is always canceled
but the pseudo-source cannot be 
canceled even if four pairs are used. This is because the angles subtended by the source 
at different pairs become different. To demonstrate this in Figs. 4(a-b) we show the reconstructed 
images of the source kept at $\phi=3000$ arcsec, $\theta=45^o$ in the source plane and at
(a) $6860$ cm and (b) $1372$ cm distance from ZP1 respectively. In Fig. 4a we see that 
the pseudo-source is present albeit very weak. In Fig. 4b, when the source is closer by, 
the pseudo-source is very strong as the positive and negative effects from the 
complimentary pairs did not cancel at all. Only when $d\sim 2$ cm, the pseudo-source
becomes weak enough to be negligible when $z=1372$ cm.

\begin{figure}[h]
\hspace{1.0cm}
\includegraphics[height=2.0in,width=2.0in]{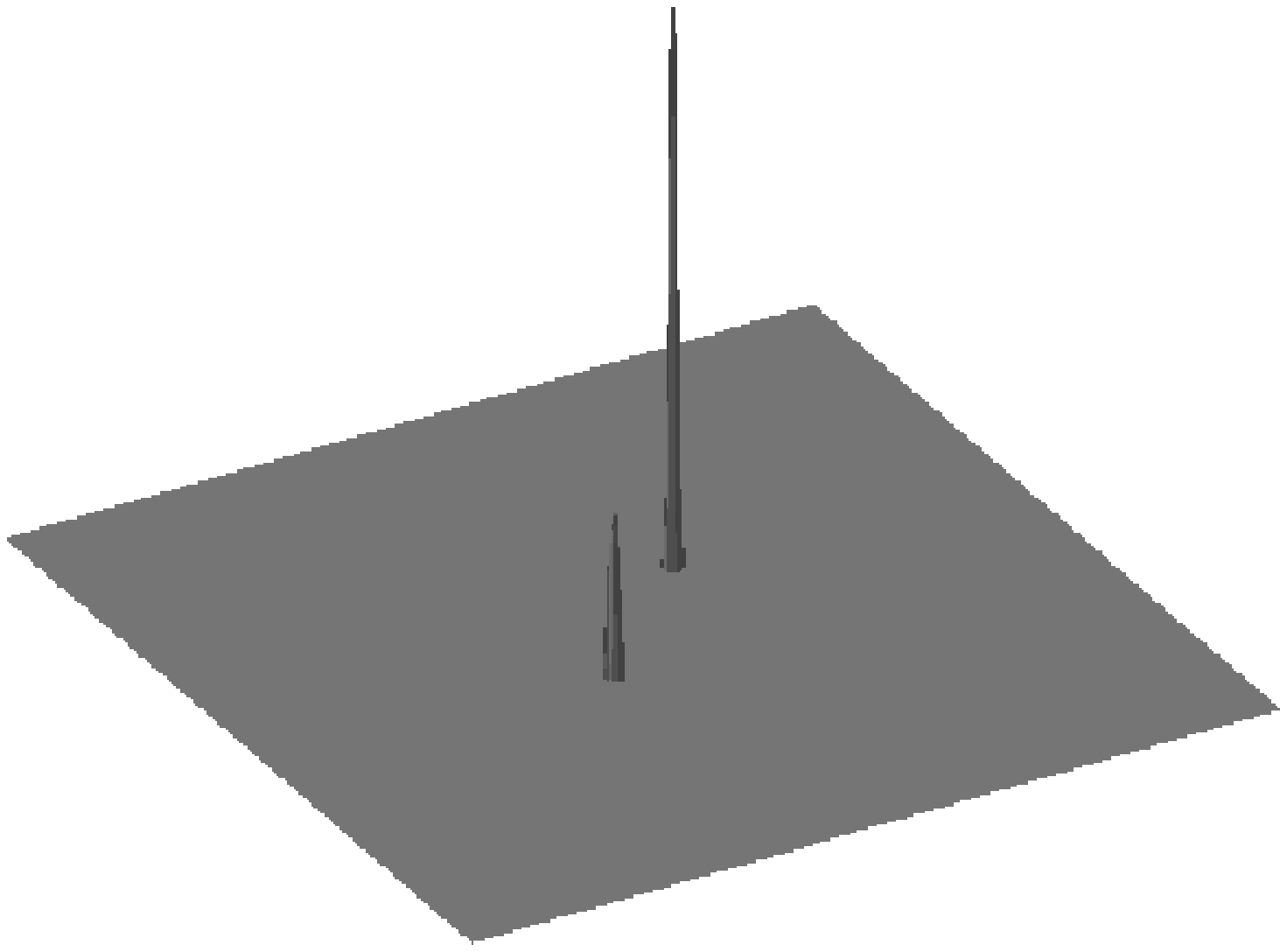}
\includegraphics[height=2.0in,width=2.0in]{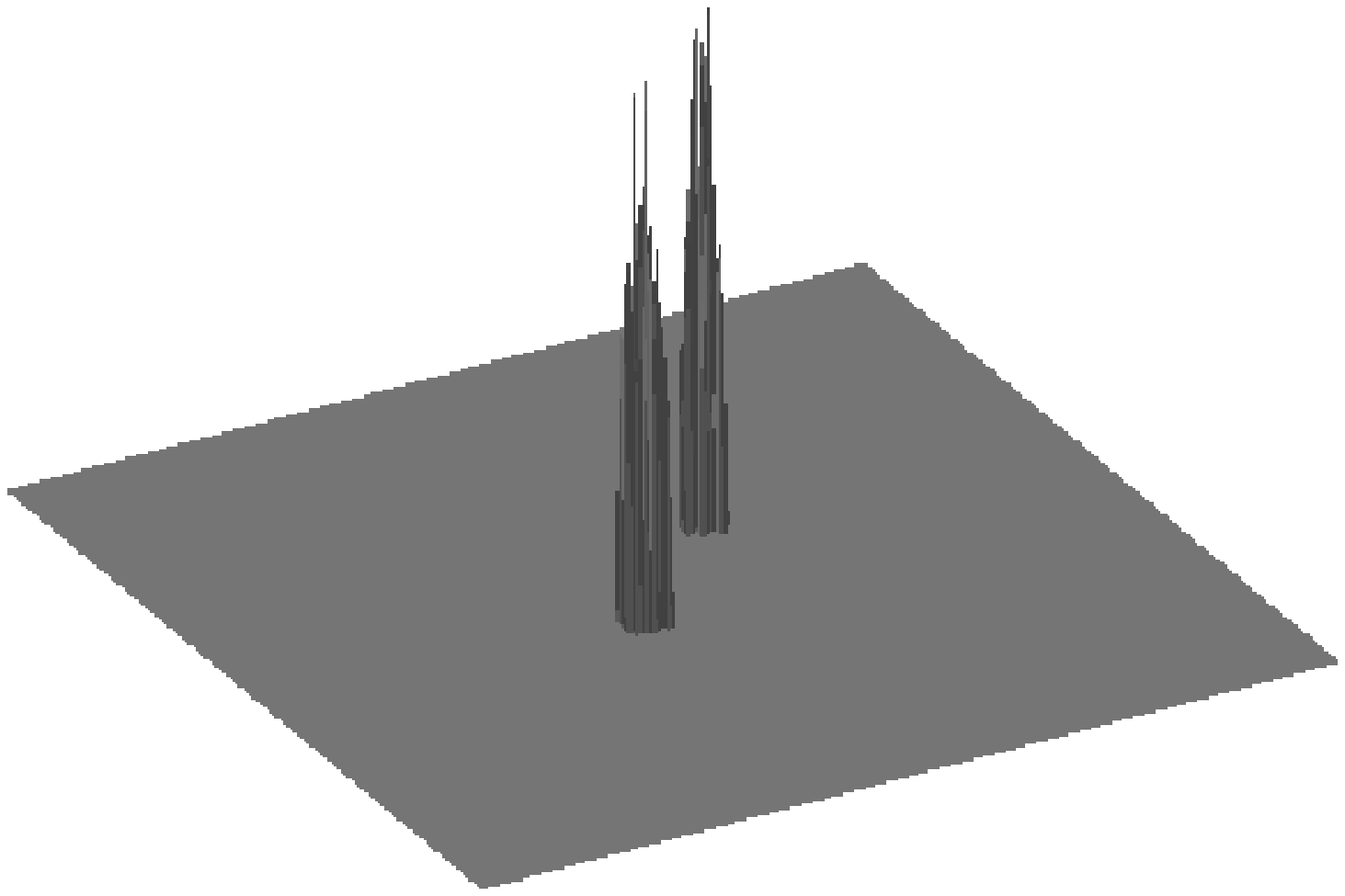}
\caption{Reconstructed images of the source kept at (a) $6860$ cm and (b) $1372$ cm
respectively as obtained by four pairs of zone plates. 
The field of view is $6.7^o \times 6.7^o$. In (a) the pseudo-source is weaker, but in (b) it is
strong. This non-cancellation of the pseudo-source is due to very close distance of the 
source from the zone plates.}
\end{figure}

\clearpage
\noindent

\subsubsection{Simulations with two pairs of zone plates}

In the last Section, we used four pairs of zone plates. From the theoretical discussion it is clear 
that if we use two pairs of zone plates, one of which gives a cosine transform and other gives  a sine 
transform, the pseudo-source will be removed though the DC-offset will remain. In an experiment, 
one has to obtain the images simultaneously and superpose the arrays and carry 
out the inverse Fourier transformation in order to reconstruct the image.

\begin{figure}[h]
\centering
\includegraphics[height=2.5in,width=2.5in]{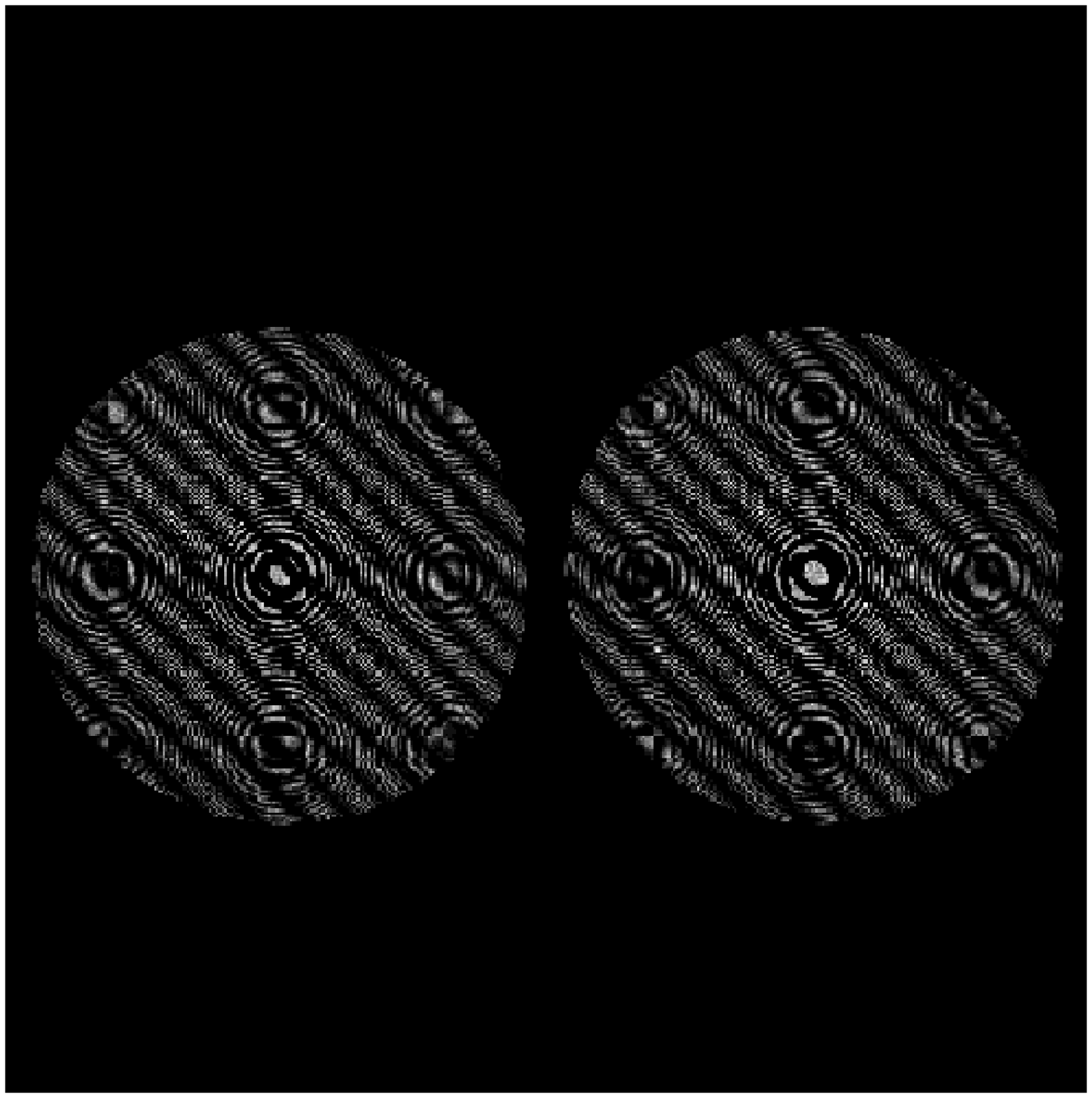}
\includegraphics[height=2.15in,width=2.15in]{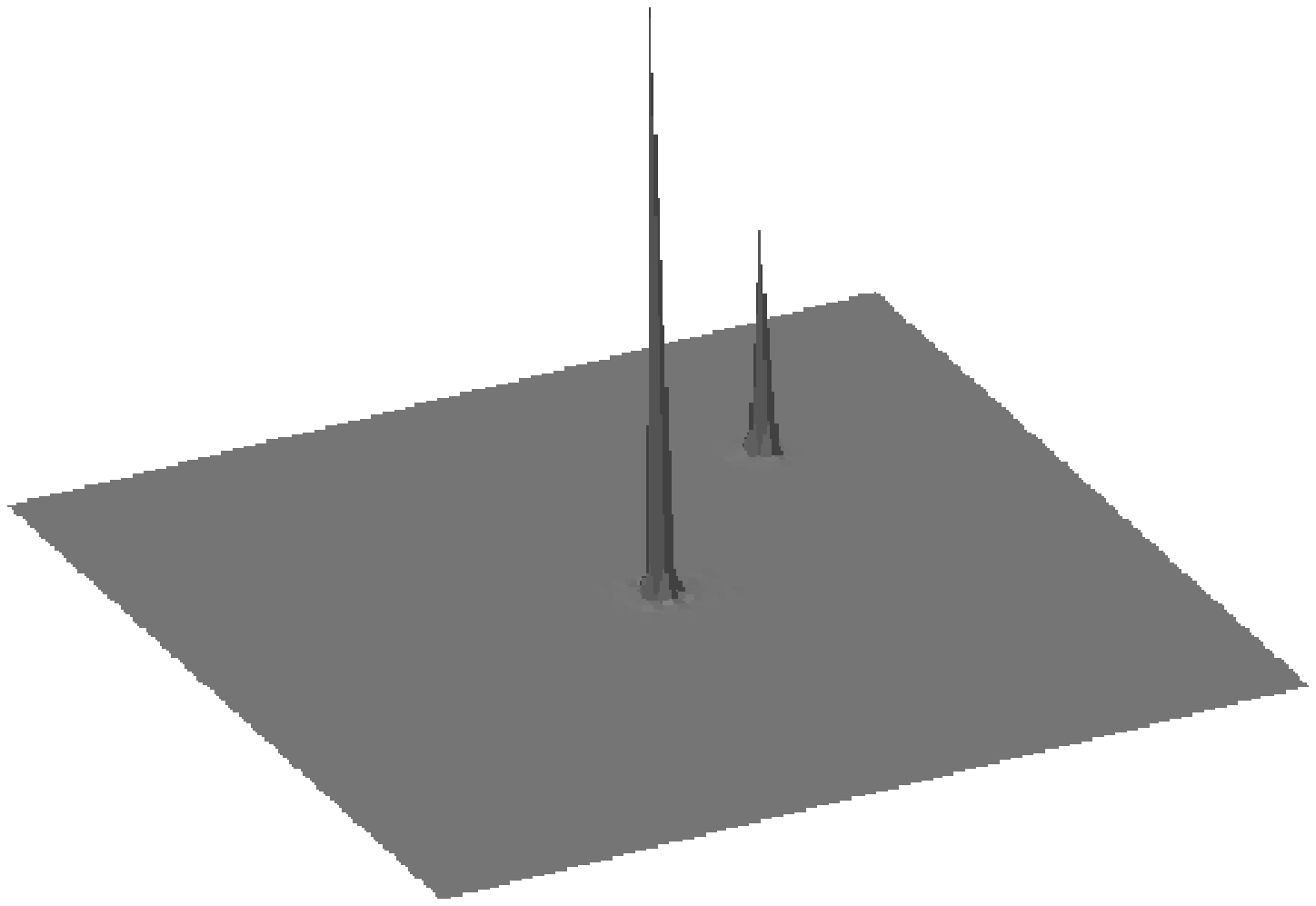}
\caption{ (a) Fringe pattern obtained with combined (sine and cosine) zone plates when source is at infinity.
(b) Reconstructed source with the DC-offset. The square is around $4600$ arcsec on each side. The
pseudo-source is totally absent.}
\end{figure}

To form the cosine pattern let us take two zone plates which are positive 
and the radii of both the plates are given by $ r_{n}  =\sqrt(n)r_{in}$, where, as before,
$r_{n}$ is the radius of the $n$-th ring and $r_{in}$ is central zone radius. Of course,
both plates having $r_{n}=\sqrt(n+1/2)~r_{in}$ or $r_{n}=\sqrt(n-1/2)r_{in}$ could also be used. 
To form a sine pattern, let us take two zone plates which are both positive, 
and the radius of $n$th zone of one plate is $r_{n} = \sqrt(n)r_{in}$ and the other plate is 
given by, $r_{n}  =\sqrt(n+1/2)r_{in}$. For the second zone plate, one could 
also use $r_{n}=\sqrt(n-1/2)r_{in}$.

In Fig. 5a we show the fringe patterns formed by two pairs of zone plates when a point source 
is placed at infinity. In the simulation, the separation between the zone plates is taken to be 
$20$ cm. The source location in its plane has the coordinates of $\phi= 1500$ arcsec and 
$ \theta= 45$ degree. The number of photons on each of the front zone plates is taken to be $10^5$. The 
reconstructed three dimensional source is presented in Fig. 5b. We observe that the DC-offset is 
still present but the pseudo-source is removed successfully.

When the source is brought nearer to the zone plate telescope, then, as in the case with four pairs 
of zone plates (\S 3.1.1), the pseudo-source reappears and becomes prominent as the distance is reduced. 
This is because the angles subtended by the source with the optical axis of the individual pairs are 
different. In Figs. 6(a-b), we present the reconstructed images. The zone plate separation is
$10$ cm and the source has the coordinate $\phi=3000$ arcsec and 
$\theta= 0$ deg. In Fig. 6a, the source is 
at $z=98$ m and in Fig. 6b, the source is at $z=14$m respectively. We see that as the source distance
is increased, the pseudo-source is becoming lesser intense as it should be.

\begin{figure}[h]
\includegraphics[height=1.8in,width=1.8in]{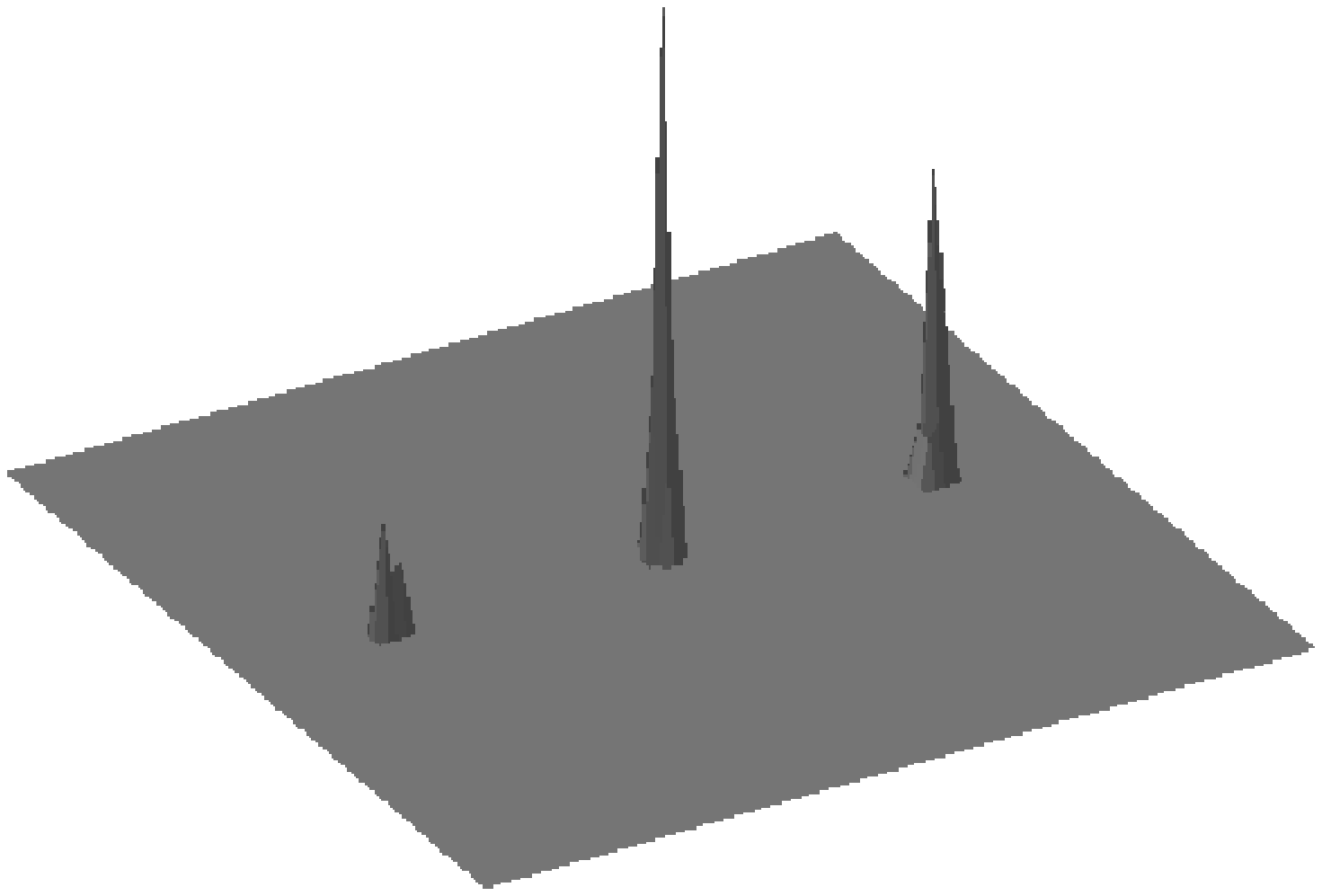}\hspace{2.8 cm}
\includegraphics[height=1.8in,width=1.8in]{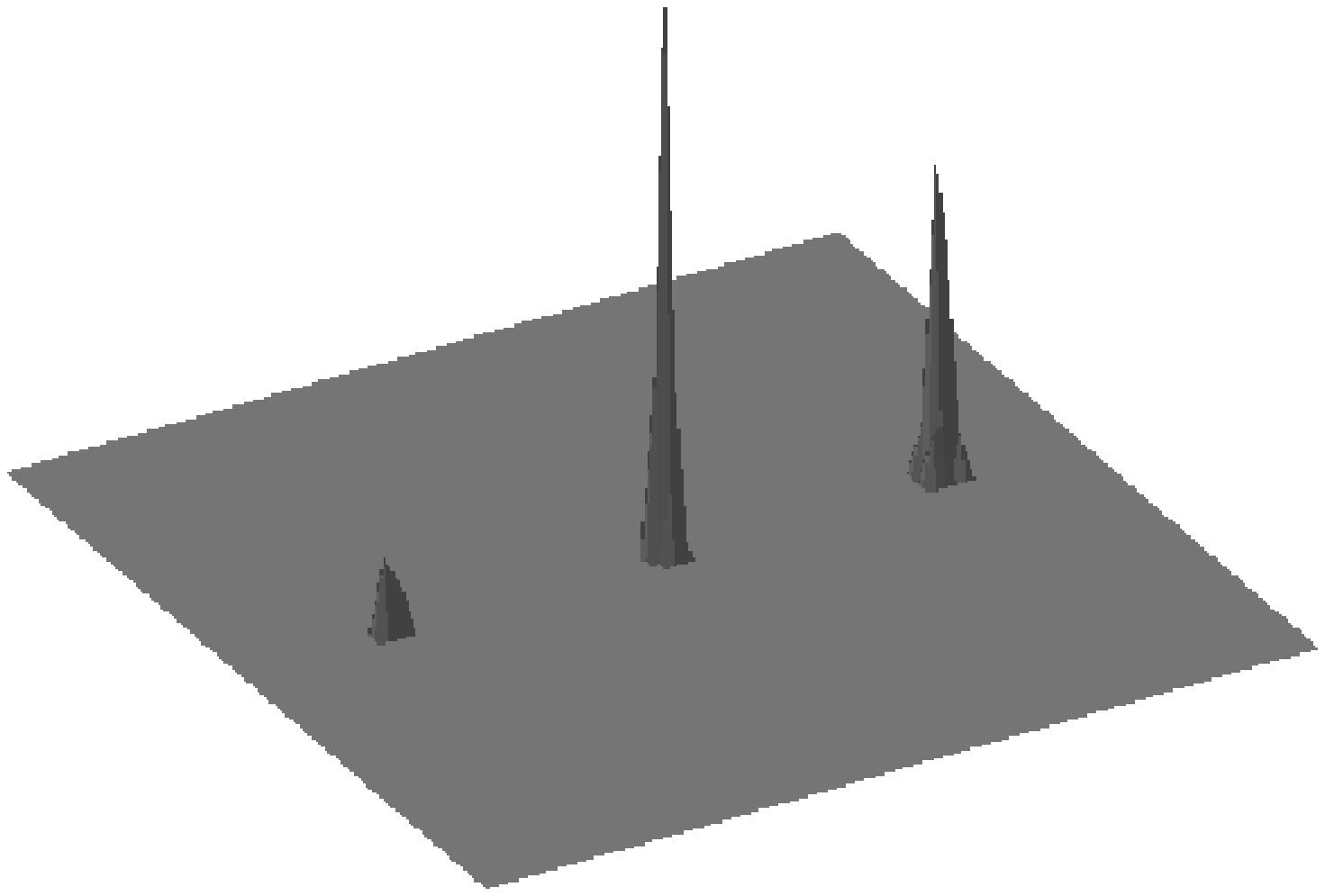}
\caption{Reconstructed image of a source at a distance of (a) $98$ m and (b) $14$m respectively.
The pseudo-source becomes stronger as the source is brought closer to the zone plate telescope. 
The image is $9000$ arcsec along one side and the number
of infalling photons on each of the zone plate pair is $10^5$.}
\end{figure}

\subsection{Resolution of the zone plate telescope}

In Paper I, it was pointed out that in astronomical imaging,
the zone plates can have arbitrarily high resolution if the plate separation is high enough.
In a realistic case for a laboratory ZPT set, the resolution is, however, moderate. The quality
of the image depends on the distance of the source from ZP1, the first zone plate. Here we systematically
show how the resolution varies with telescope parameters. For clarity of the image, 
we use four pairs of zone plates in each of these simulations.

\subsubsection{Sources at infinity}

From Eq. (1), we note that for zone plates with a spacing of $40$ cm between them, the resolution should be $51.5$ arcsec.
In Fig. 7(a-b), we present the reconstructed images of a pair of sources which were placed at an angular 
distance of (a) $103$ arcsec and (b) $51.5$ arcsec (limiting case). Each side of the square is $20$ arcmin 
in length.  

\begin{figure}[ht]
\includegraphics[height=1.5in]{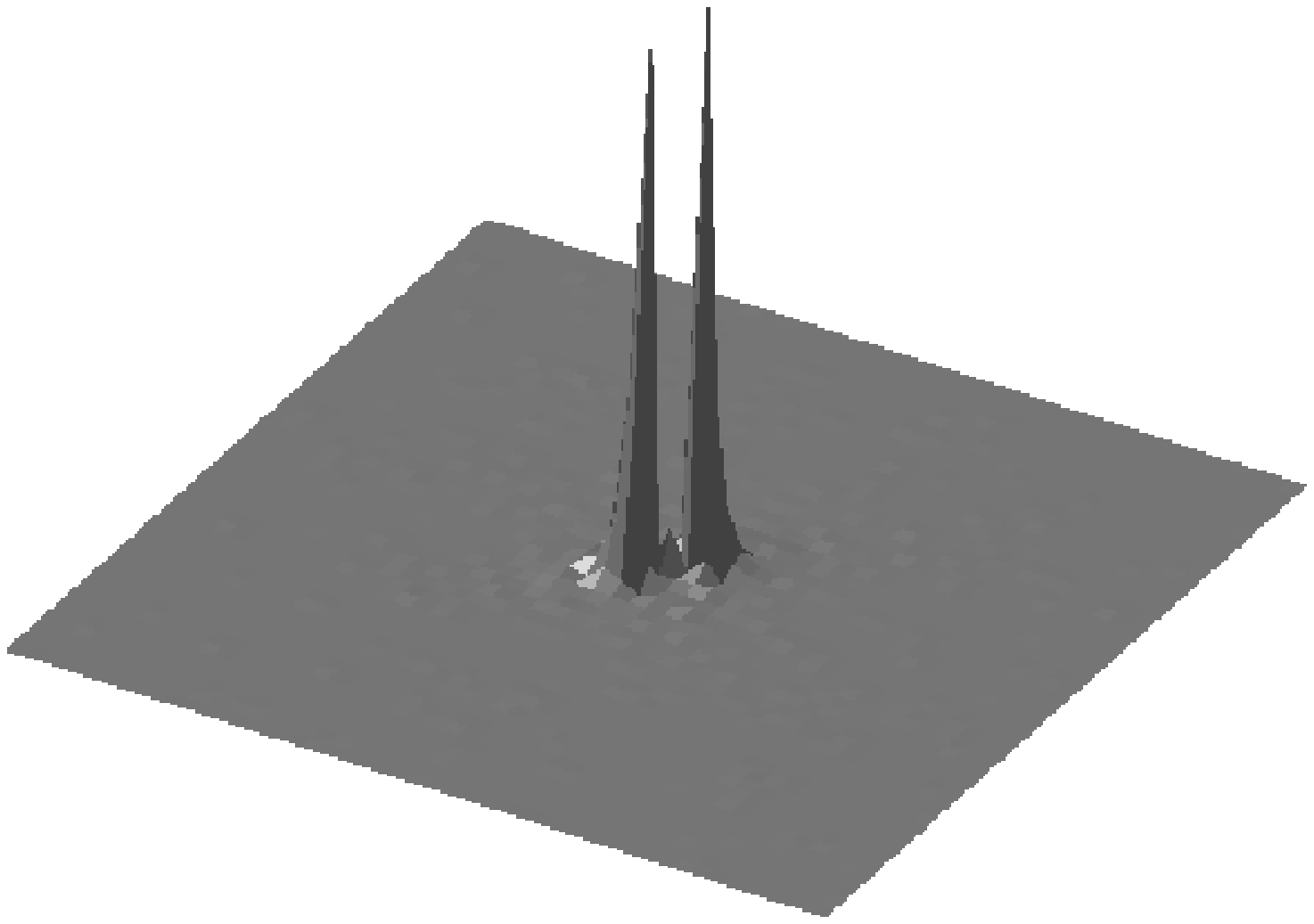} \hspace{0.5cm}
\includegraphics[height=1.5in]{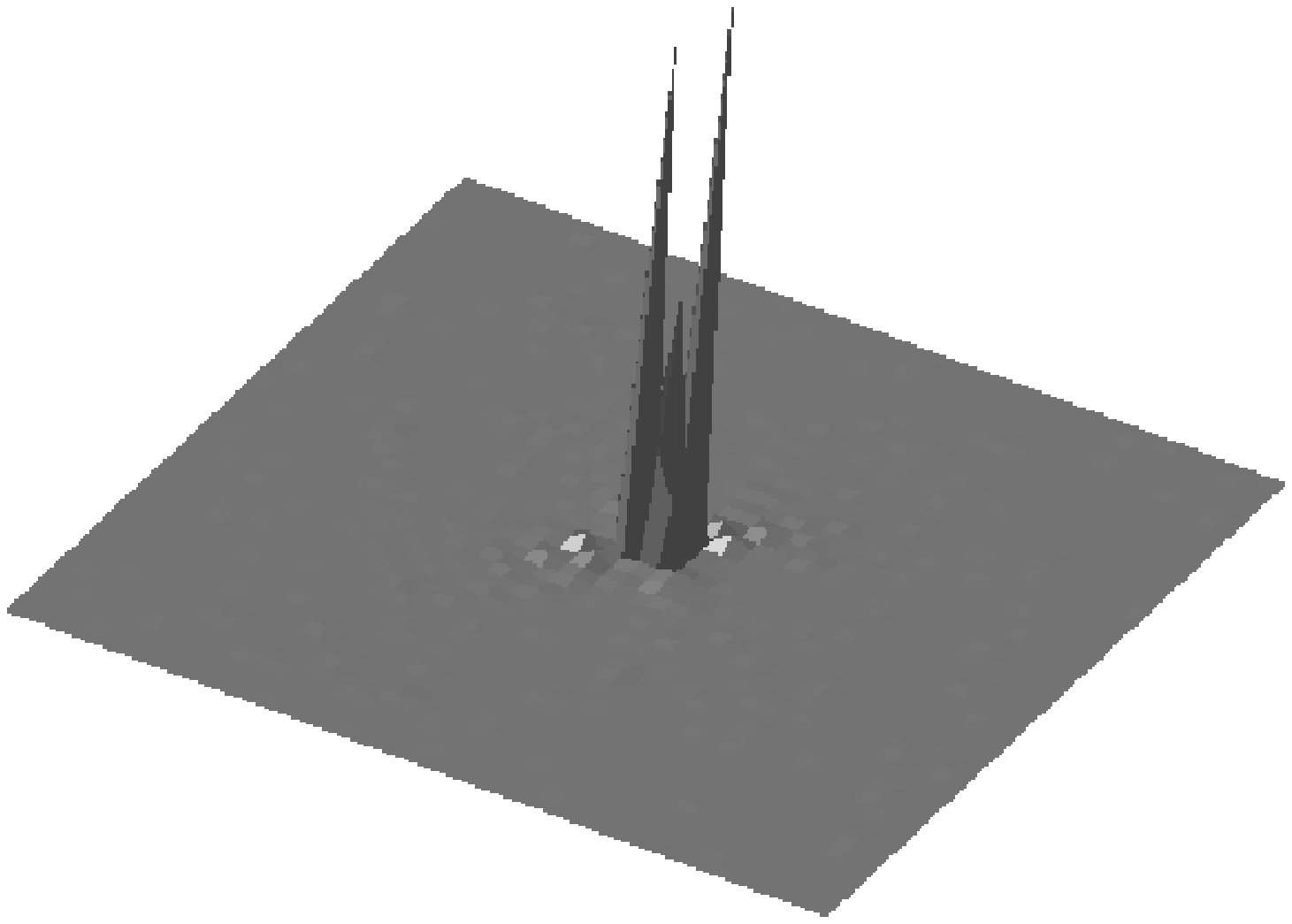}
\caption{Sources placed at infinity and separated by (a) $103$ arcsec and (b) $51.5$ 
arcsec respectively are reconstructed from the fringe patterns. The zone plate 
separation is $40$ cm. In each Figure, the base is $20$ arcmin on each side.}
\label{}
\end{figure}

Similarly, with separations of $20$ cm and $10$ cm respectively, we get the angular resolutions 
to be $103$ and $206$ arcsec respectively as predicted by the theoretical estimate.

From Eq. (1), we see that the resolution is directly proportional to the inner radius $r_{in}$ 
of the zone plate. This is because the radius of the $n$th zone is proportional to $r_{in}$.
In order to show this, we take one of the cases we simulated above. With $d=20$ cm and 
$r_{in}=0.122$ cm, $\theta_r =103$ arcsec. However, for $r_{in}=0.183$ cm, the sources separated by
$103$ arcsec should not be resolved as is clear from Fig. 8a. However, when the 
angular separation is increased to $\theta_r=154.5$ arcsec, they are just resolved (Fig. 8b).
Thus the dependency on $r_{in}$ is also verified. 

\begin{figure}[h]
\includegraphics[height=1.5in]{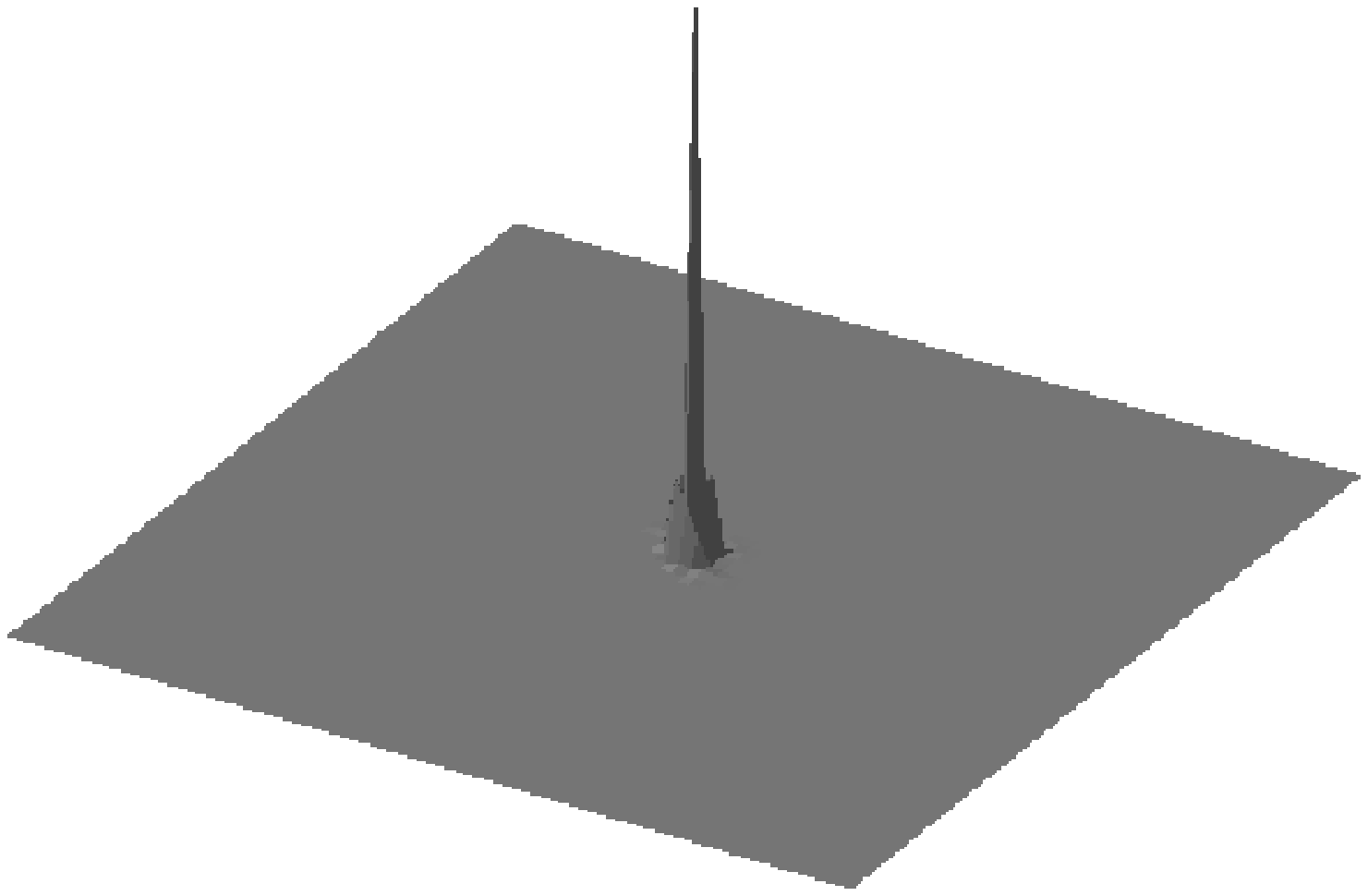} \hspace{0.5cm}
\includegraphics[height=1.5in]{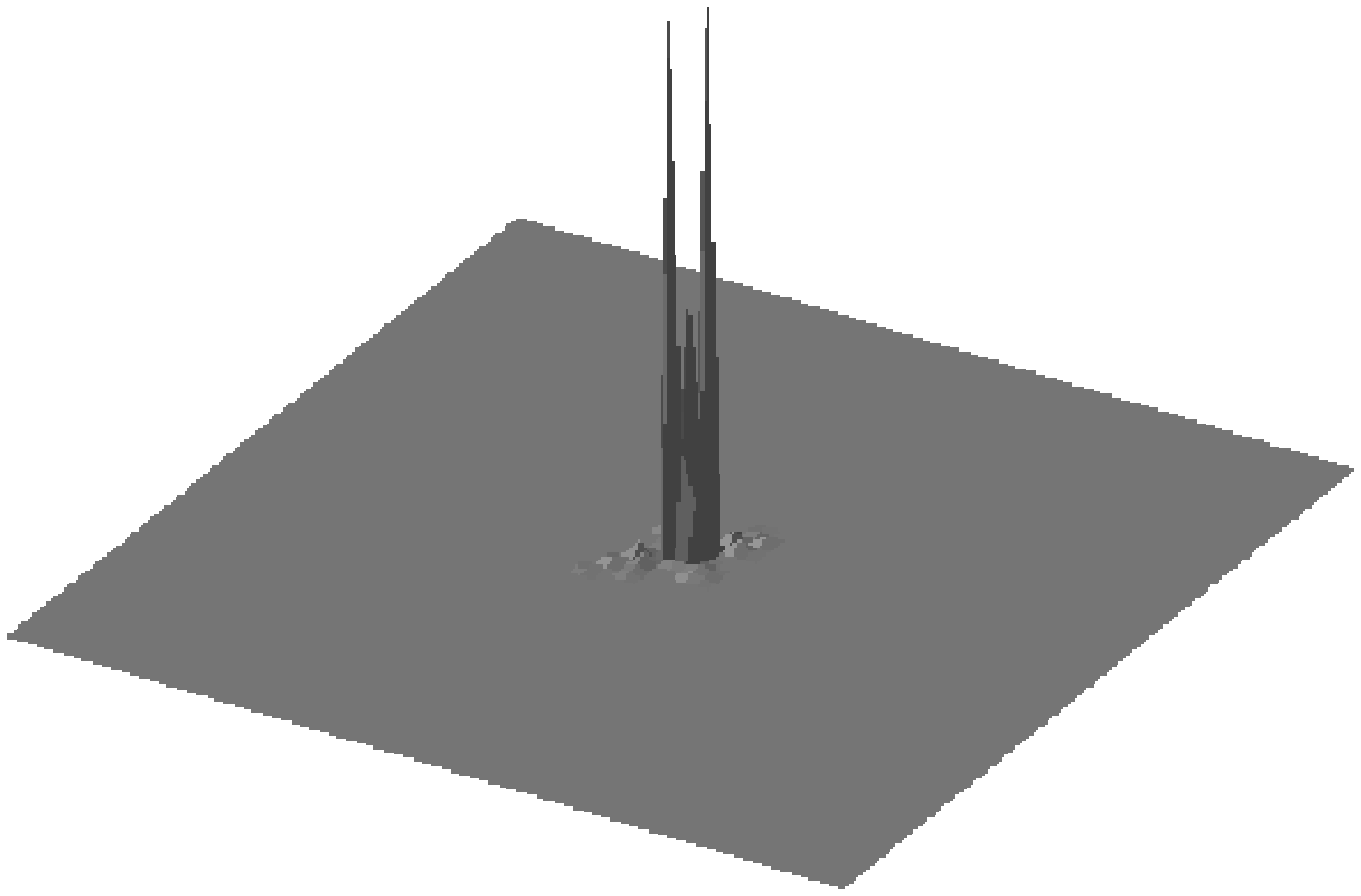}
\caption{Sources placed at infinity are separated by (a) $103$ arcsec and (b) $154.5$ arcsec respectively. 
Here $d=20$ cm and $r_{in}=0.183$ cm. The squared base is $3900$ arcsec along each side. In (a), 
sources are not separated but in (b), they are just separated, as expected.}
\label{}
\end{figure}

\subsubsection{Sources at a finite distance}

In \S 2.2, we discussed how the reconstructed image of a point source 
is broadened due to its finite distance from the zone plate telescope.
To show this effect, we carry out the following simulation with two sources.
One is placed at $\phi = 2400$ arcsec (angular distance from the ZPT axis)
and $\theta = 0$ deg (angle measured counter-clockwise from the X-axis) and the other 
is placed at $\phi = 1200$ arcsec and $\theta = 90$ deg. The sources are placed at a 
distance of $D=1372$cm from the front zone plate (ZP1) as in the experimental 
results presented in Paper I. The zone plate separation is taken to be $d=100$ cm. 
In Fig. 9a, we show the fringe pattern and in Fig. 9b, we show the reconstructed image
($5800$ arcsec on each side). It is clear that location wise, the sources have been placed properly, 
although they look `similar' to the Moir\'{e} fringe patterns, which have special shapes due to 
off-axisness of the sources. In the simulations, the number of photons impinging from these 
two sources are taken to be $10^5$ and $7\times 10^4$ respectively, causing one pattern to be 
slightly brighter than the other. The image size is $\sim 550$ arcsec.
This is somewhat higher than the expected value of $\sim 420$ arcsec due to the fact that there is 
a further spreading due to finite size of the CMOS detector pixel, each of which is $50$ micron.

\begin{figure}[h]
\centering
\includegraphics[height=1.9in]{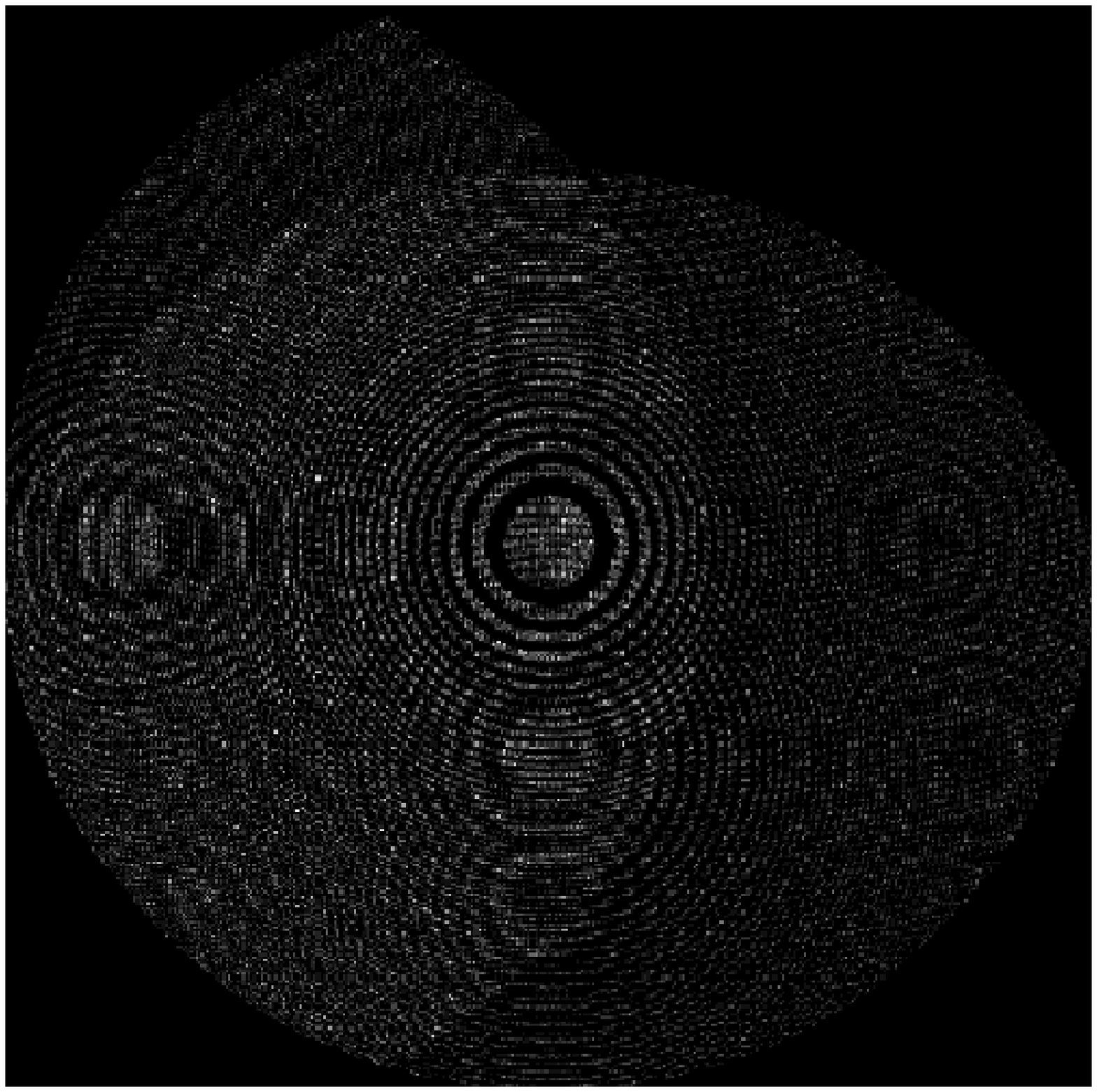} \hspace{0.5cm}
\includegraphics[height=1.9in]{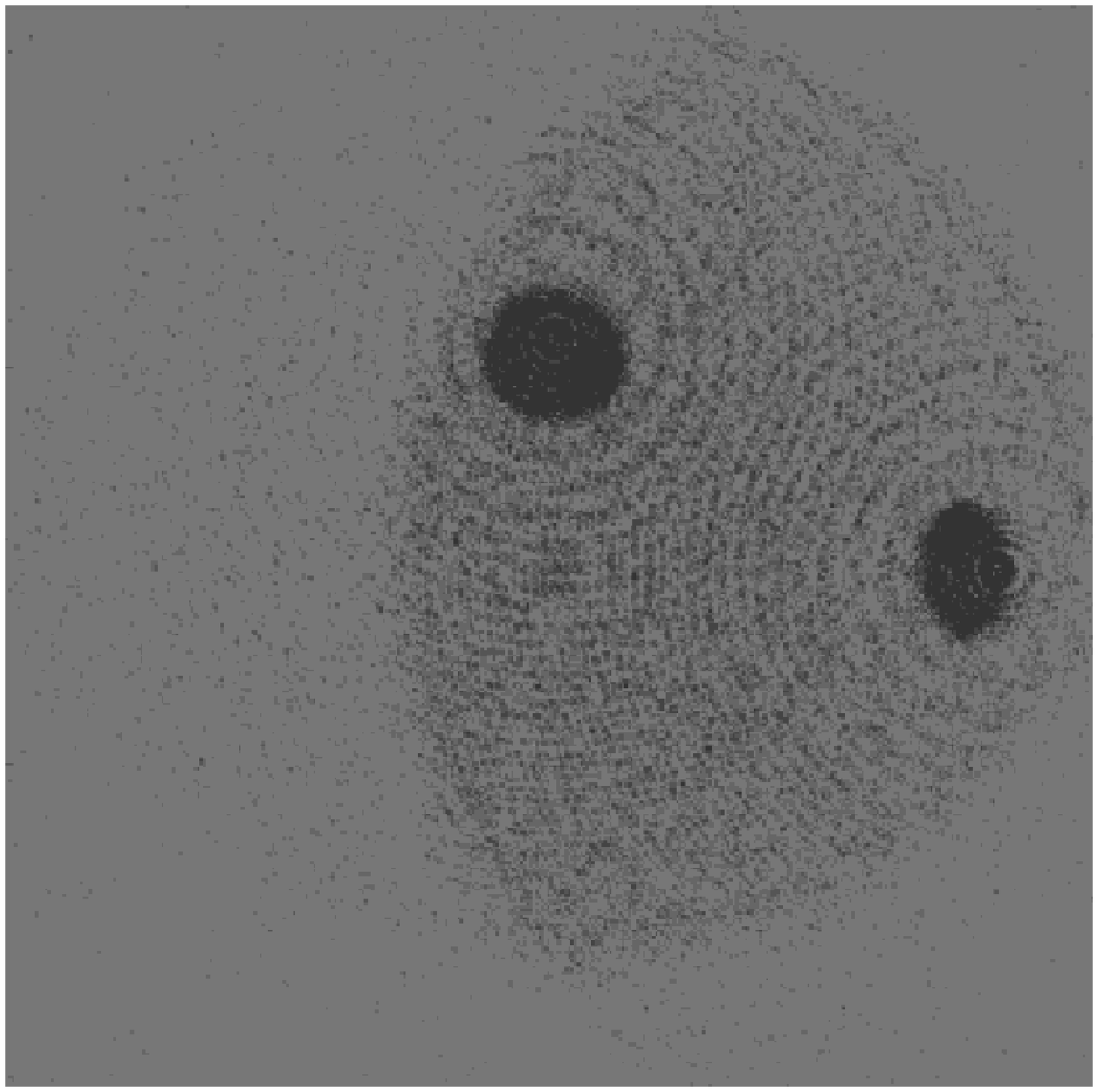}
\caption{(a) Moir\'{e} fringe patterns obtained for two sources placed at about $1372$ cm away from the 
front zone plate. In the telescope, the plates are separated by $100$ cm. (b) The reconstructed 
images. Due to large off-axisness the fringes do not cover the entire zone plates. The image size
is compatible with the distance of the source and the image shape is compatible with the 
degree of off-axisness.}
\label{}
\end{figure}

In \S 2, we showed that the resolution becomes inferior if the source is close to the 
zone plate (see, Eq. 3). The reason being that the reconstructed image of a point source 
placed at a finite distance acquires a finite angular size. With increasing $d$, the 
zone plate separation, the inter-source distance in the reconstructed image 
increases. This causes a better resolution for the sources placed at infinity as there is no effect 
of broadening. However, for the sources at a finite distance, the  broadening increases further with the 
zone plate separation ($d$). So sources may even overlap for sufficient broadening and resolution 
may worsen with an increment of $d$. As an example, we show in Fig. 10(a-b) a distribution of $51$ 
point sources in the shape of ICSP. Two successive sources are at an angular distance of $300$ arcsec.
The zone plates are separated at a distance of $20$ cm. In Fig. 10a, the sources are kept at a large, 
effectively infinite distance and the point sources are clearly separable. 
In Fig. 10b, the sources are placed at a distance of $1372$cm,
same as in our experimental setup (Paper I). Each point source spreaded, but the image is recognizable.
Here each side of the square is $8500$ arcsec long.

\begin{figure}[h]
\centering
\includegraphics[height=1.9in,width=1.9in]{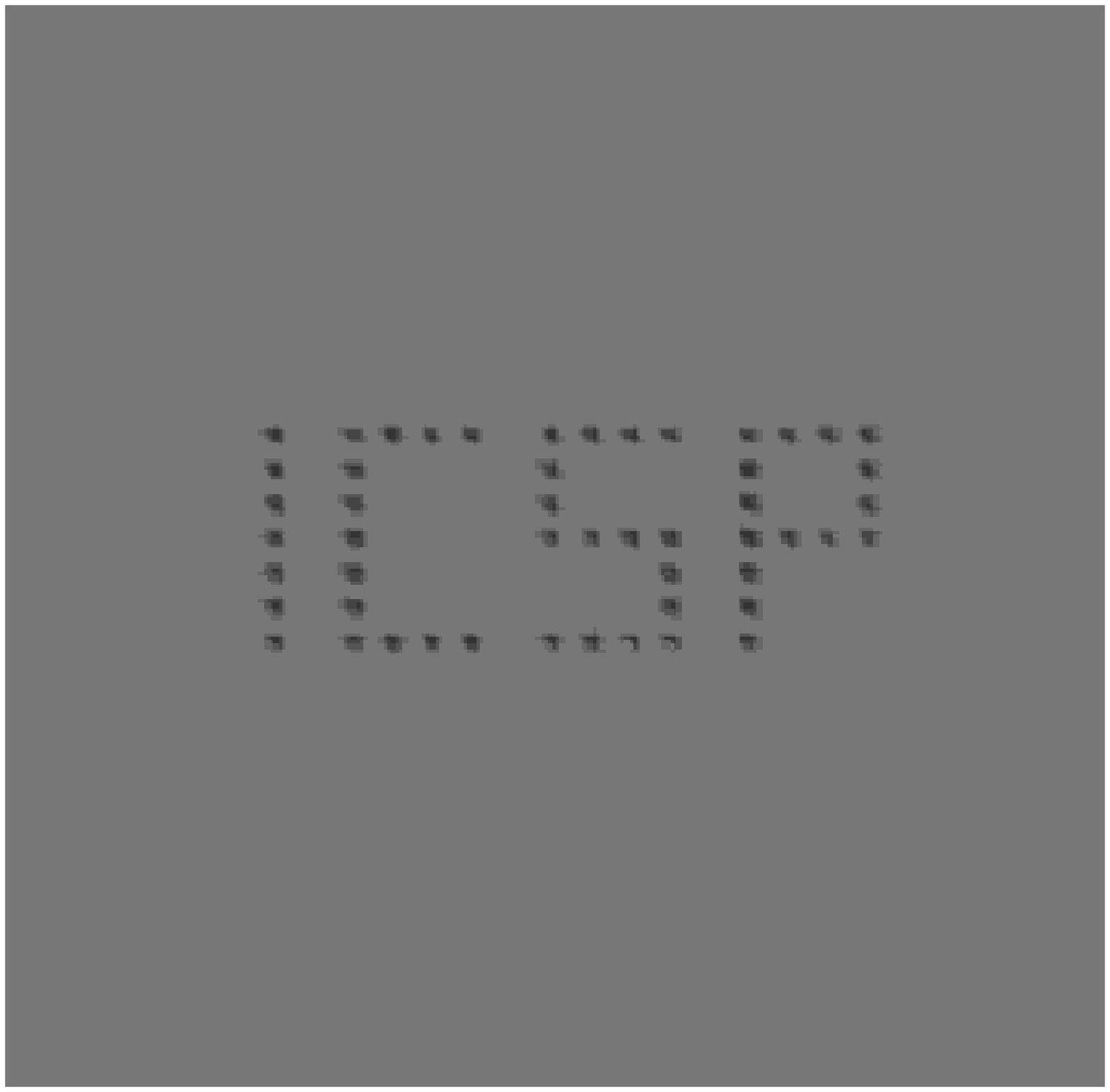}\hspace{0.5 cm}
\includegraphics[height=1.9in,width=1.9in]{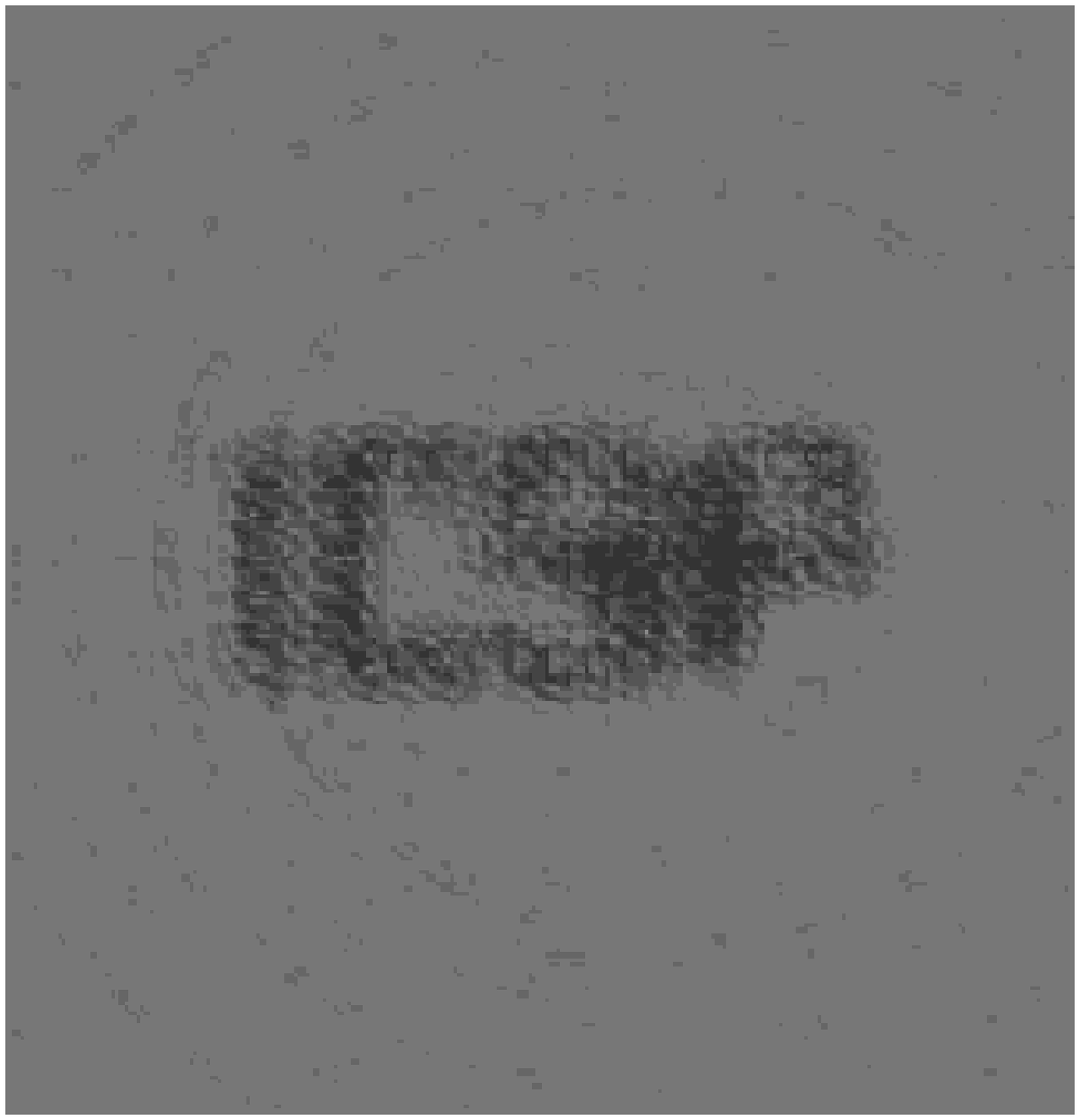}
\caption{A distributed source resembling ICSP at large distance reconstructed with a zone plate
telescope having $20$cm plate separation. (a) The source distance is very high and (b) the source
is at a distance of $1372$ cm (see text for details).}
\end{figure}

\newpage
In Fig. 11, we show the reconstructed image of the same case 
with zone plate separation increased to $40$ cm. Though the distance 
between two successive images is higher, the resolution is worse, since 
the source distance should have been twice as big to get the same resolution (Eq. 3). Here,
the longer side is $5800$ arcsec long.

\begin{figure}[h]
\begin{center}
\includegraphics[height=2.0in,width=2.0in]{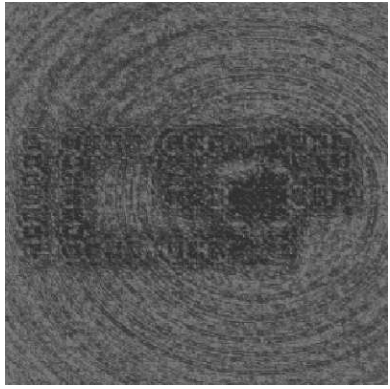}
\caption{Same as in Fig. 10(b), but the plate separation is $40$ cm. Image is blurred.}
\end{center}
\end{figure}

\subsection{Resolution with modified zone plate telescope}

In \S2 we have pointed out that for obtaining high resolution in case the sources at a
finite distance the angular size of the zone plates as subtended at the source must be the
same. In Figs.  12(a-b) we show the results presented in Figs. 9b and 10b using a modified 
zone plate telescope where such a consideration has been implemented. We clearly see that the 
images have becomes sharp, as though the source is at infinity. 

\begin{figure}[h]
\centering{
\includegraphics[height=2.0in,width=2.0in]{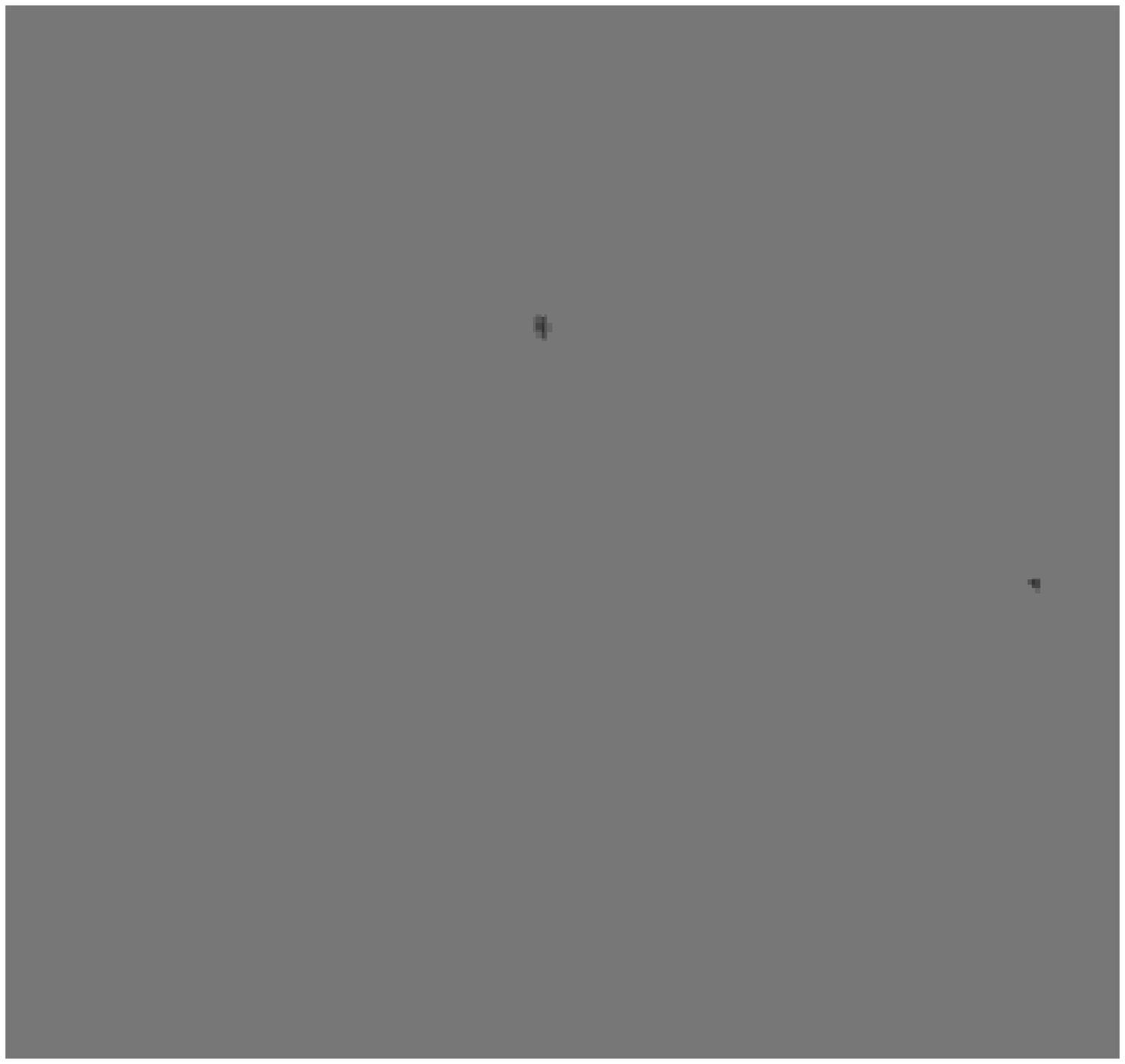}
\hspace{0.2 cm}
\includegraphics[height=2.0in,width=2.0in]{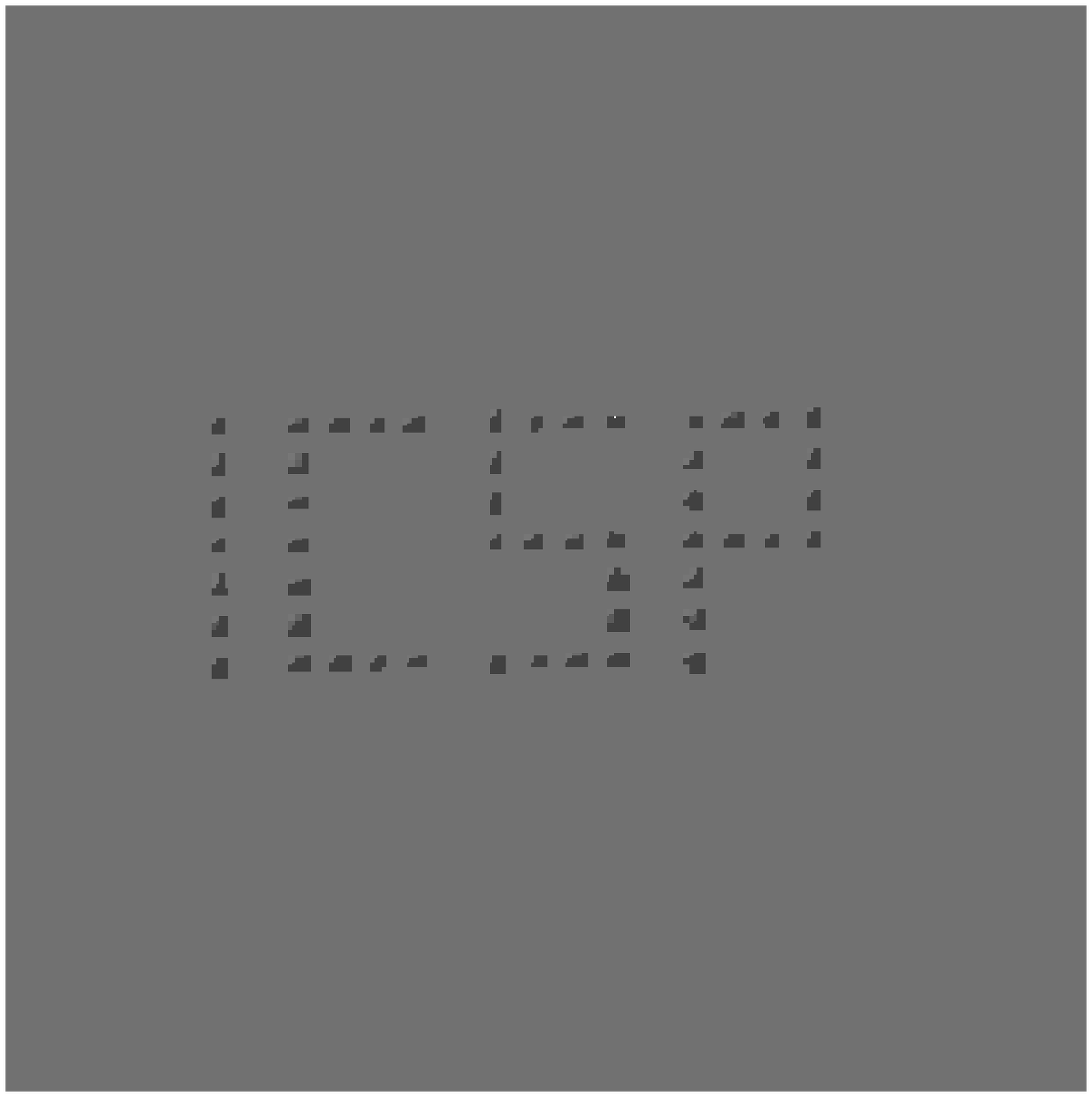}}
\caption{(a) Both sources as presented in Fig. 9b are seen through a 
modified ZPT. The images appear to be sharp. (b) The  distributed source as considered in Fig. 10b 
observed using a modified ZPT. The image is very sharp as expected.}
\end{figure}

\subsection{Dependence on source intensity}

One of the advantages of carrying out numerical simulations is that one can vary 
the source parameters at will. In a zone plate with $151$ zones, we have $75$ transparent zones. 
In order to find how many photons are needed to reconstruct the image uniquely, 
we note that to pinpoint the centre of the plate, we need at least three photons to pass through
a single zone. The rest of the $74$ zones may have one photon each. So, on an average 
if $77$ photons enter through the plate, then the 
imaging could be done with all the zone plane information taken into account. Of course,
lower number of photons will skip a few transparent zones and the image would be noisy.
Note that our argument does not depend on the overall area of the zone plate, but only depends 
on the total number of zones. So with a larger zone plate having the same number of zones
will detect fainter sources.

In Figs. 13(a-b) we demonstrate how the photon number affects the quality of the 
reconstructed image. In Fig. 13a, $1000$ photons are allowed to fall on each of the 
front zone plates and the image is sharp.
However, when only $100$ photons are allowed to fall, the image is reconstructed, but it is noisy.

\begin{figure}[h]
\centering
\includegraphics[height=1.9in,width=1.9in]{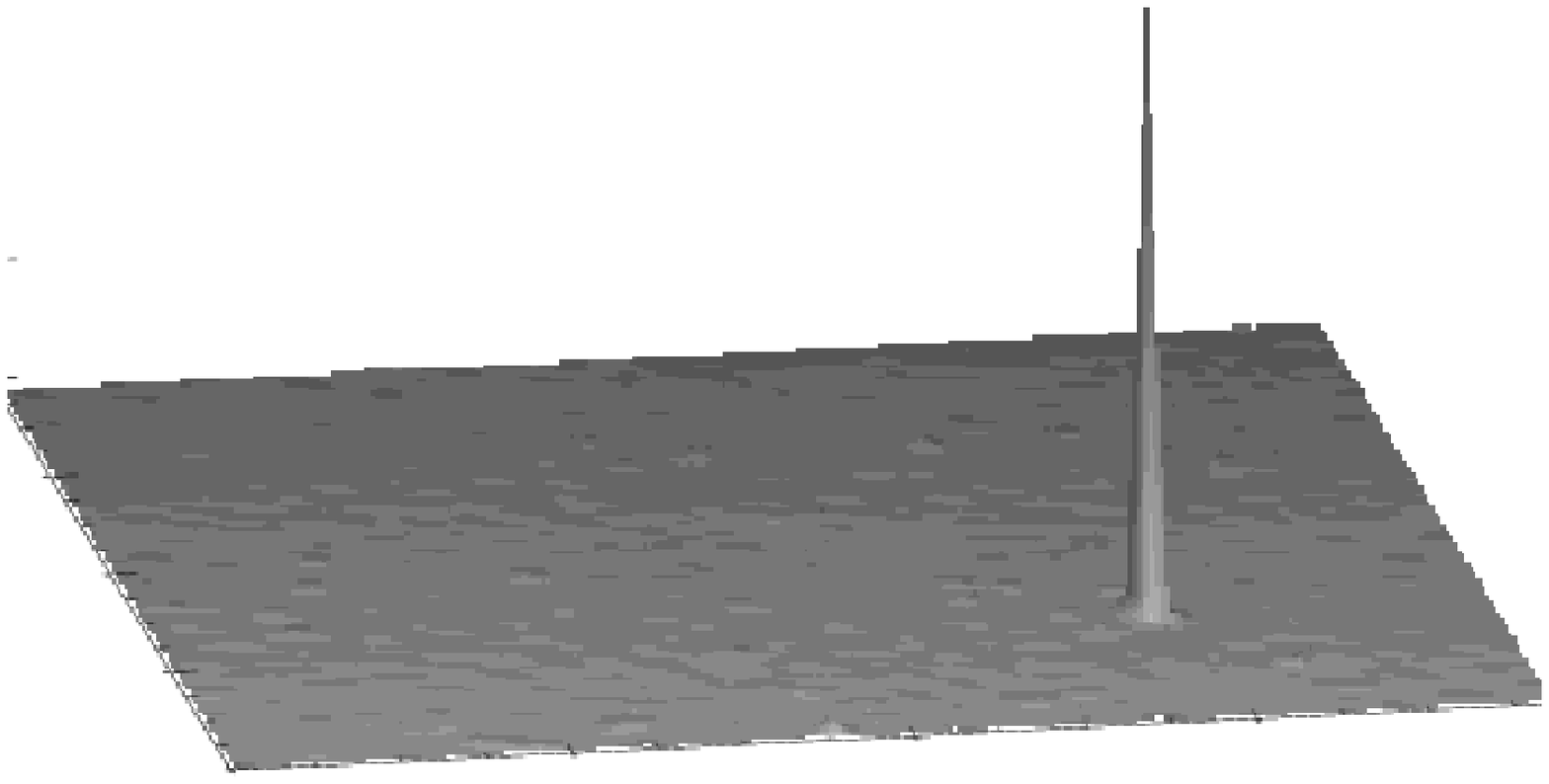} \hspace{0.0cm}
\includegraphics[height=1.9in,width=1.9in]{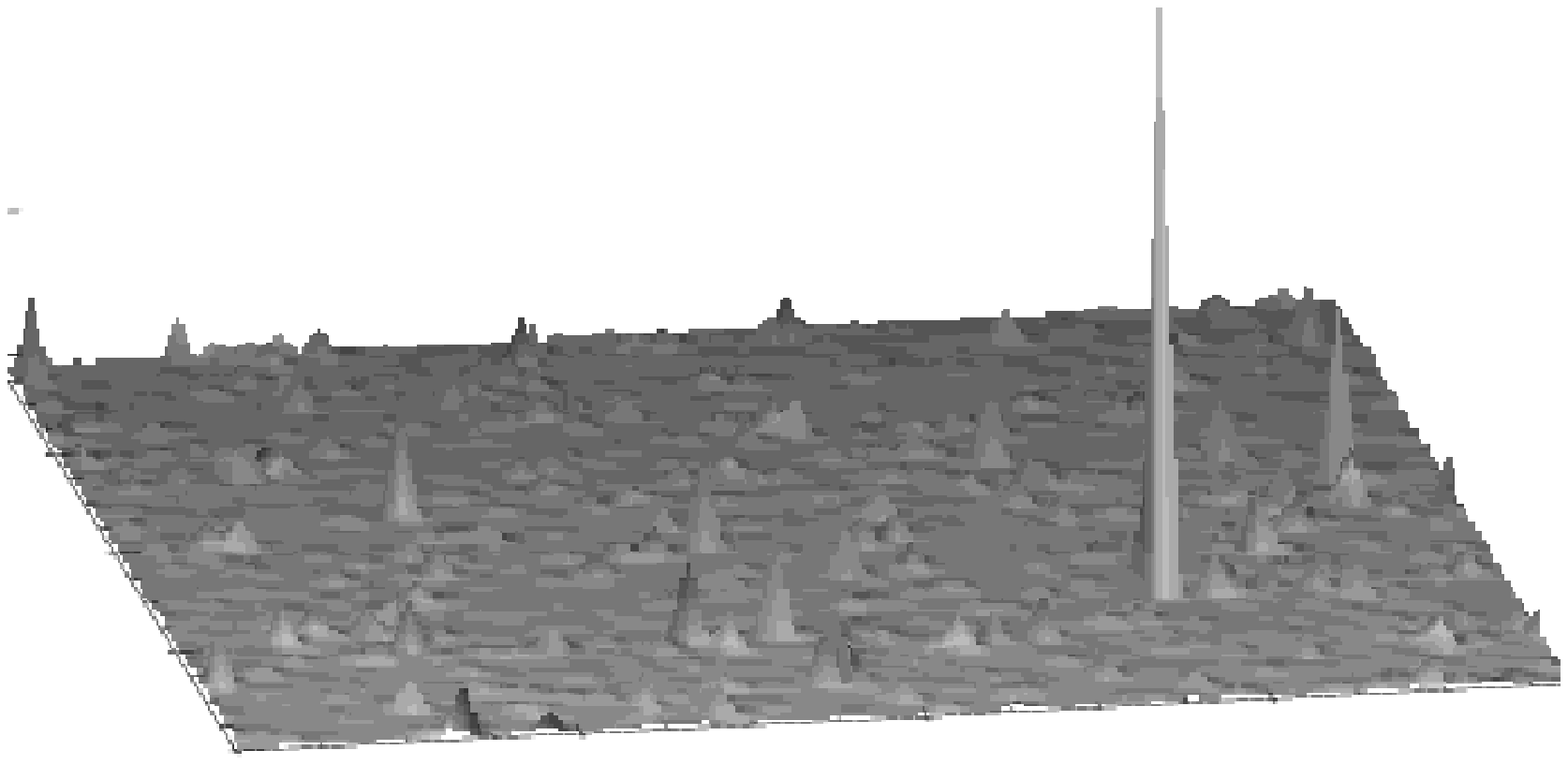}
\caption{Reconstructed image of a point source at infinity by four pairs 
of zone plates with (a) $1000$ photons, and (b) $100$ photons.}
\end{figure}

\subsection{Simulation of X-ray Sources near the Galactic center}

We now give an example of how to reconstruct images in a situation where the 
intensity varies from source to source. Our Galactic center has 
several relatively bright X-ray sources which can be imaged by a ZPT. We
start with four pairs of zone plates as before. The simulation is made with 
zone plate spacing of $20$ cm. The source distribution and the intensity 
variation have been obtained from INTGCCAT, the INTEGRAL IBIS Hard X-ray 
Survey of Galactic Center. Here, the X ray sources are detected 
in the energy range 18-60 keV. In the two-dimensional Figures the 
vertical direction is the direction of celestial north-south and
horizontal direction is the direction of celestial east-west. 

In Fig. 14, we plot the 3D image of how the Galactic center should look 
like. The following sources have been considered (parenthesis gives the 
intensity in mcrab): SLX 1744-299/300 (7.6), 1E1740.7-2924 (74.8), 
A 1742-294 (10.4), KS 1741-293 (2.1), AX J1745.6-291 (57.4),
1E1742.8-2853 (6.2), IGR J17475-2822 (2.2), SLX 1737-282 (3.4). 

\begin{figure}[h]
\centering
\includegraphics[height=2.5in,width=3.0in]{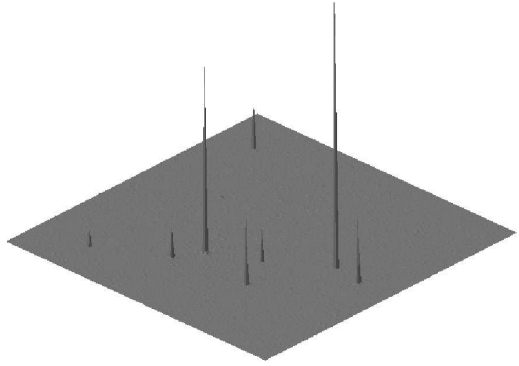}
\caption{3D plot showing eight sources with accurate relative photon 
counts near the Galactic center. The field is $3.5^o \times 3.5^o$.}
\end{figure}

We now present the result of the simulation in which we examine how the fringe system should look like
when such a system of sources are turned on. In Fig. 15a, we show the fringe patterns 
produced by the four pairs of zone plates on a CMOS detector ($50$ micron pixel size). 
The plate separation is $20$ cm. The reconstructed source distribution is seen in Fig. 15b.

\begin{figure}[h]
\centering
\includegraphics[height=2.3in,width=2.3in]{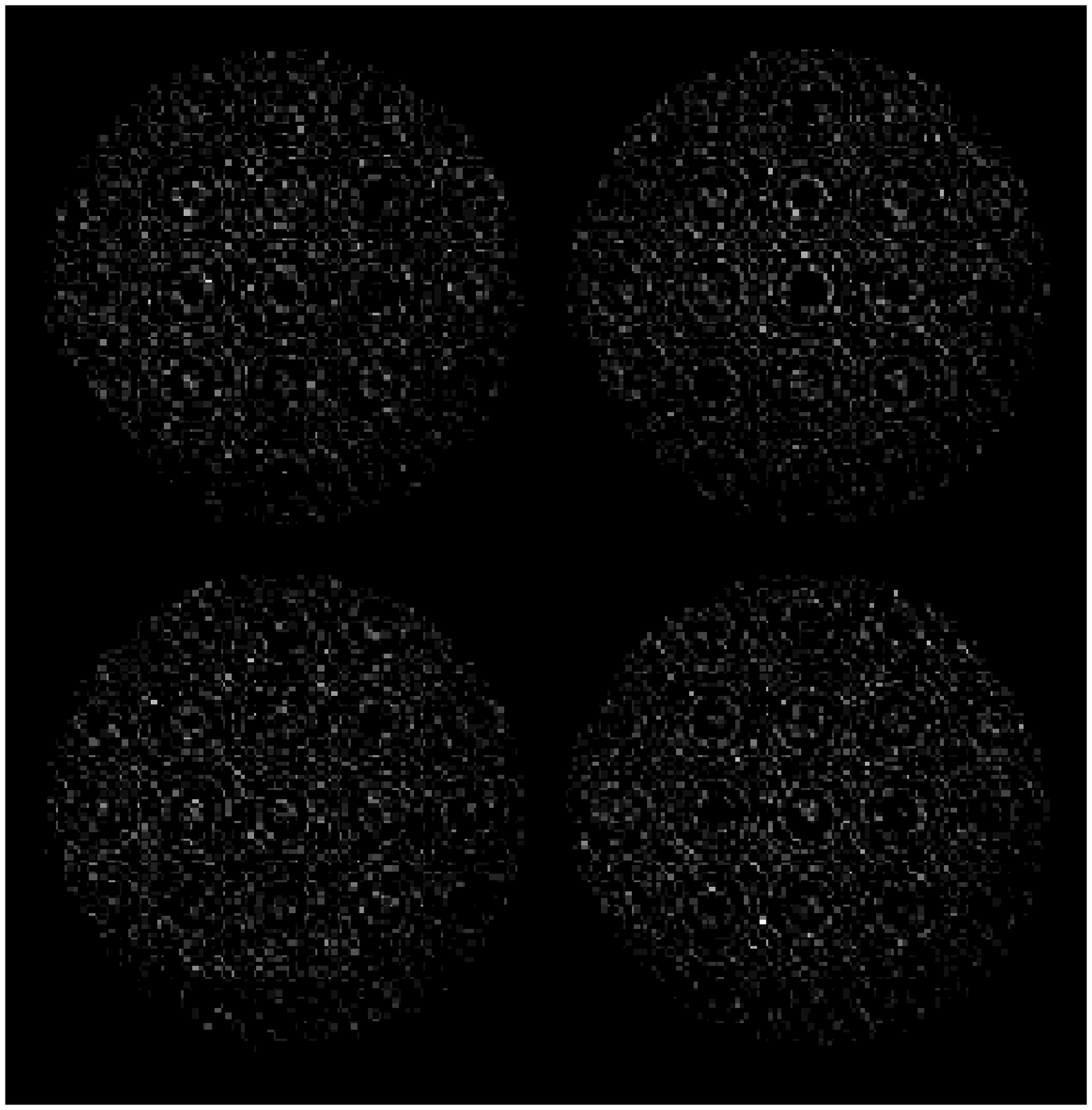}
\includegraphics[height=2.2in,width=2.4in]{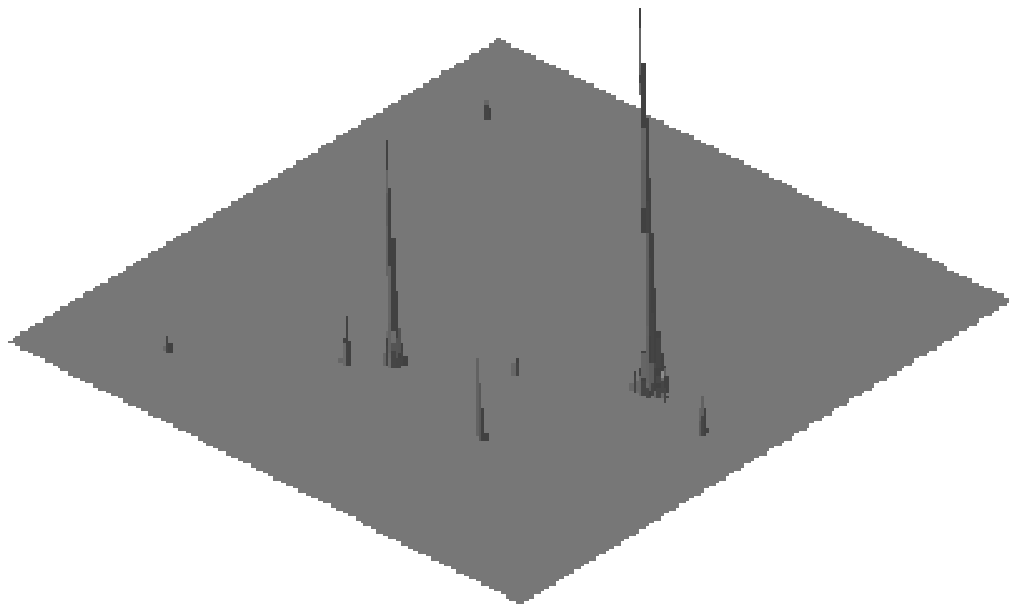}
\caption{(a) The fringe patterns with four pairs of zone plates for actual X-ray source 
distributions near the Galactic center. (b) The reconstructed source distribution from the fringes. 
The field of view is $3.5^o \times 3.5^o$.}
\end{figure}

\newpage
A better way to produce the images of the source having an intensity distribution would be first to 
reconstruct the sources and subtract the fringes produced by the most intense sources
from the original fringe system. 

\begin{figure}[h]
\centering
\includegraphics[height=2.3in,width=2.3in]{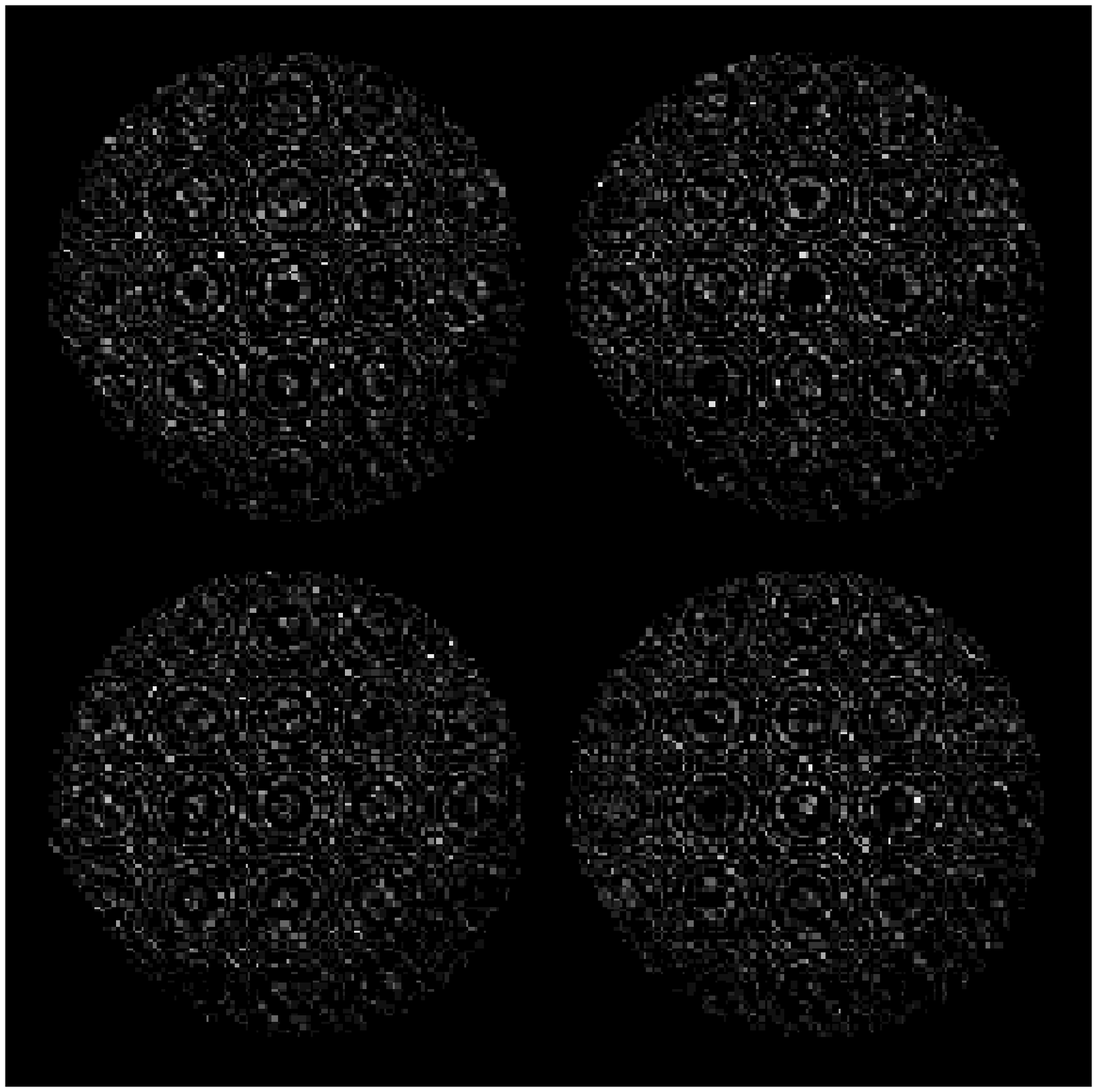}
\includegraphics[height=2.2in,width=2.2in]{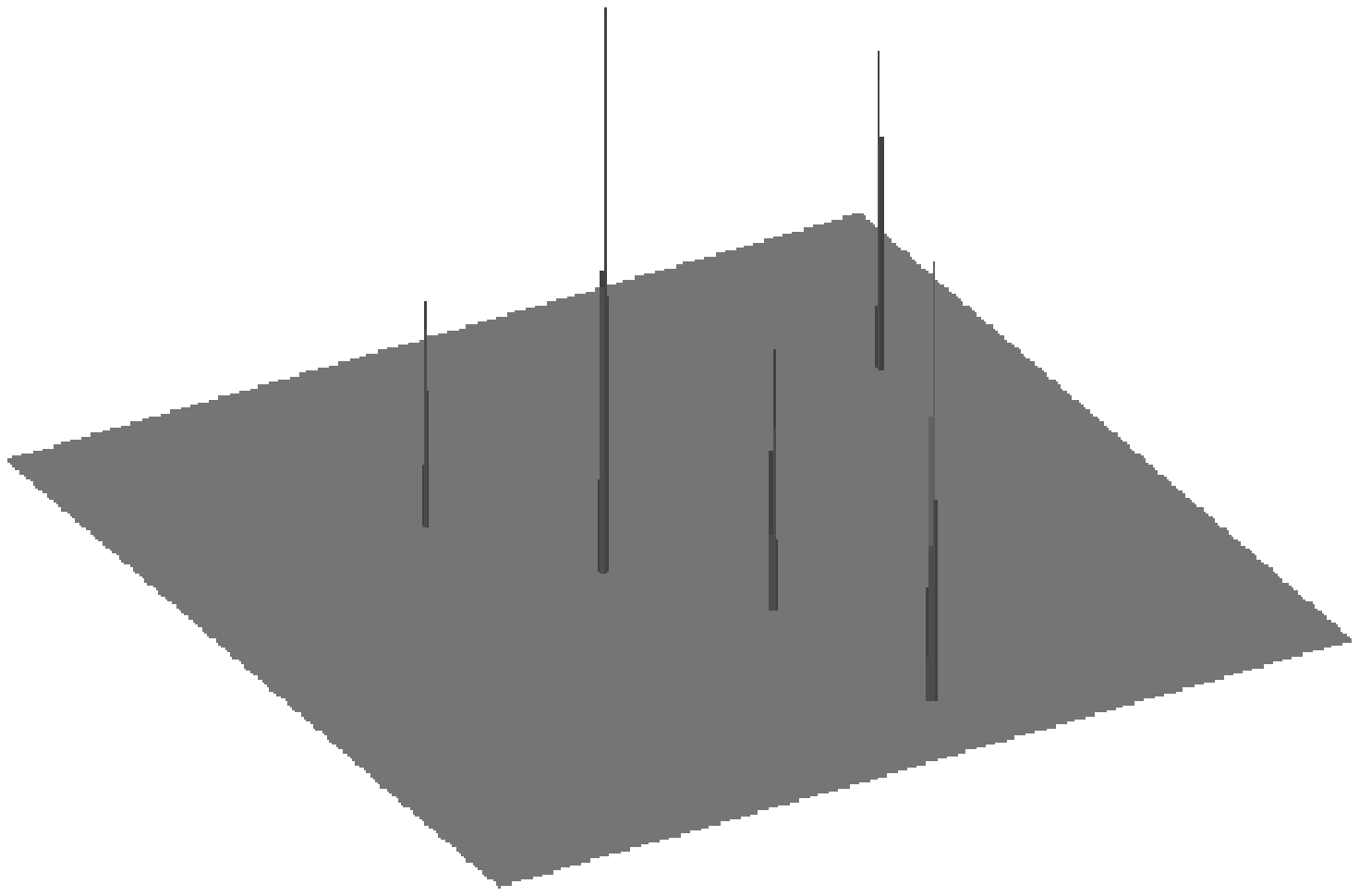}
\caption{(a) Fringes obtained with four pairs of zone plates for three most intensed sources.
(b) The reconstructed sources from the fringe pattern obtained by subtracting the fringe
pattern of (a) from those presented in Fig. 15a.}
\end{figure}

In this particular case, we construct the fringe system
of three most intense sources such as, 1E1740.7-2942 (74.8 mcrab), AX J1745.6-2901 (57.4 mcrab)
and A1742-294 (10.4 mcrab). Then we subtract these fringes from the original one (fringes with eight 
sources), and reconstruct the images of the weaker sources. This procedure gives the 
remaining five sources distinctly. Fig. 16a gives the fringes obtained with three 
intense sources and Fig. 16b gives the 3D view of the reconstructed weaker sources 
from the subtracted fringe pattern. This procedure could be successively used to obtain 
even more weaker sources if present. The time of integration used in the 
simulations is about one hour. For larger zone plates, the time of integration would be
proportionately reduced.

\section{Simulation with Cadmium Zinc Telluride (CZT) detector}

Since we are motivated to study the images of the solar flares by RT-2 payloads aboard CORONAS-PHOTON
where both the CMOS and CZT detectors have been used, we now discuss the image reconstruction for
CZT detector. The disadvantage with a CZT detector is that the pixel size is big ($0.25$ cm), but the 
advantage is that it is possible to obtain the energy dependence of the image. In the case of 
CMOS detector, on the other hand, the pixel size is smaller ($50$ micron) and hence the 
image is of high resolution.  However, the images do not have information of 
energy and thus energy integrated images are obtained.
 
Here we present the results of the simulations with zone plates having inner radius $0.122$ cm
and number of zones $151$. The zone plate separation is $10$ cm. The large
pixel size in CZT limits the field of view. In this case, the effective field of view is 
about $409$ arcsec along each side. In Fig. 17a, we show the fringes obtained when a source is 
kept at an angular distance of $\phi=65$ arcsec from the optical axis and $\theta=45$ degree (i.e.,
along a diagonal of the zone plate. In Fig. 17b, we show the reconstructed image. The image is noisy
but recognizable.

\begin{figure}[h]
\centering
\includegraphics[height=2.0in,width=2.0in]{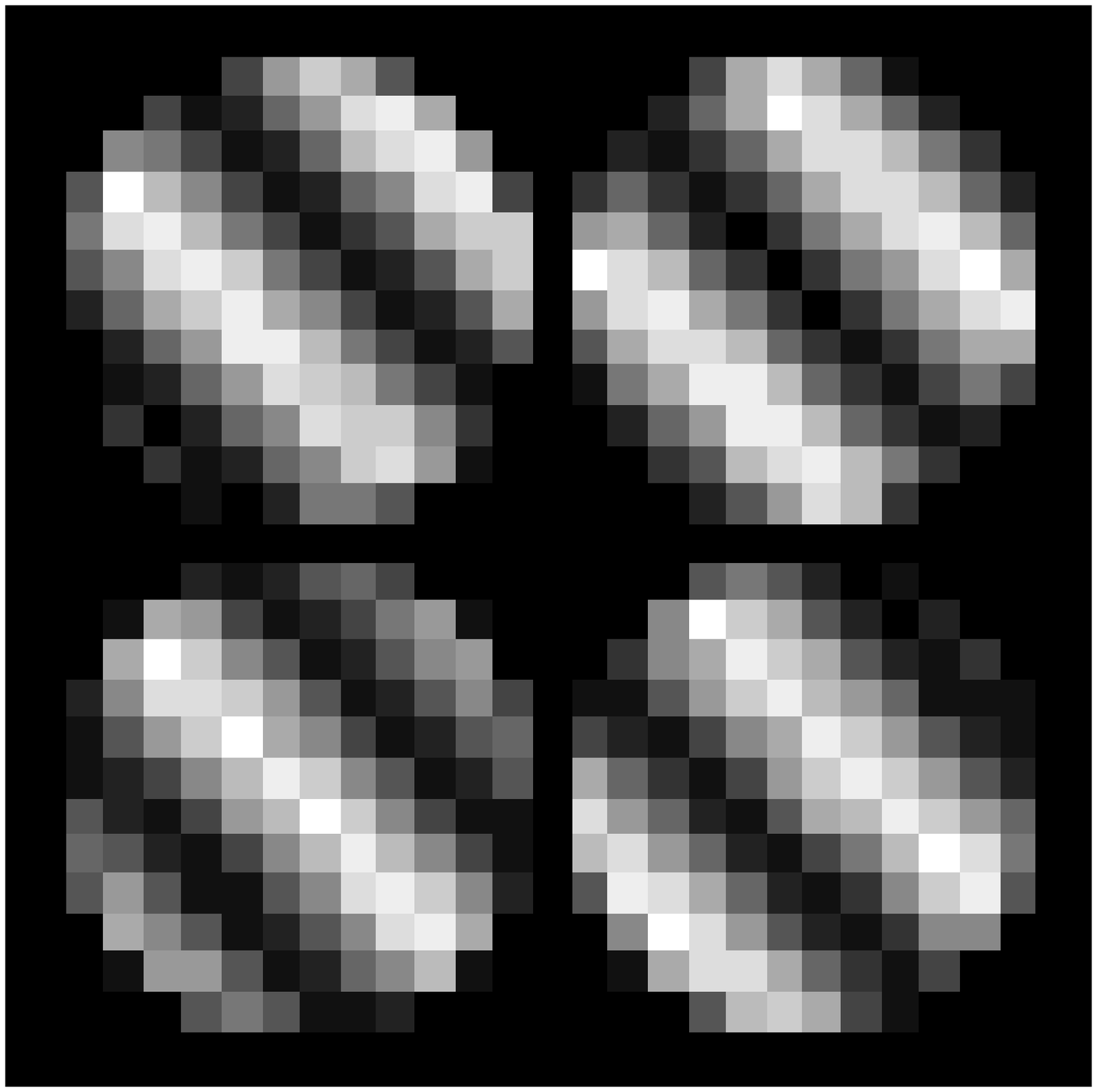}\hspace{0.2 cm}
\includegraphics[height=2.0in,width=2.0in]{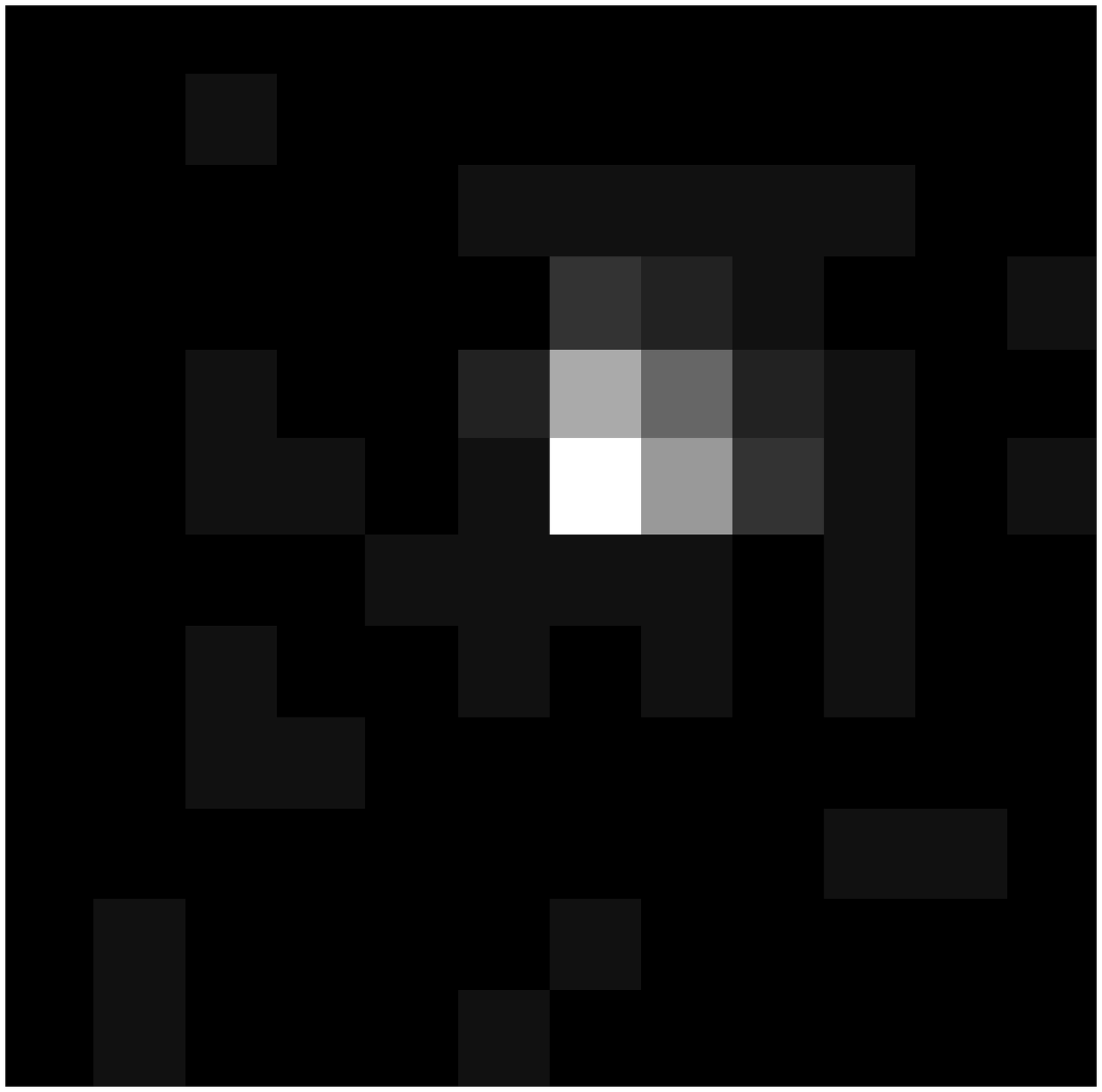}
\caption{ (a) Fringes obtained with four pairs of zone plates with a CZT detector for a source 
kept at $\theta=45$ degree at an angular distance of $65$ arcsec from the 
optical axis. (b) Reconstruction of the image from the fringe system.}
\end{figure}

In the next set of simulations, we present results when the source is placed at a maximum offset,
either horizontally or vertically and also diagonally. In this case, we take the zone plate 
separation of $30$ cm. Here the maximum offset along the vertical/horizontal direction 
($\theta = 0$ deg or $\theta = 90$ deg) is $204.5$ arcsec and along diagonal (say, $\theta = 45$ deg) 
it is equal to $289.5$ arcsec. In Fig. 18a, we present the simulated fringes on a CZT detector
for a source placed at $\phi=204.5$ arcsec and $\theta=0$ degree. In Fig. 18b, the image of the 
reconstructed source has been given. In Figs. 19(a-b), we similarly compute the fringe pattern when the 
source is placed at at $\phi=289.5$ arcsec and $\theta=45$ degree.

\begin{figure}[h]
\centering
\includegraphics[height=2.0in,width=2.0in]{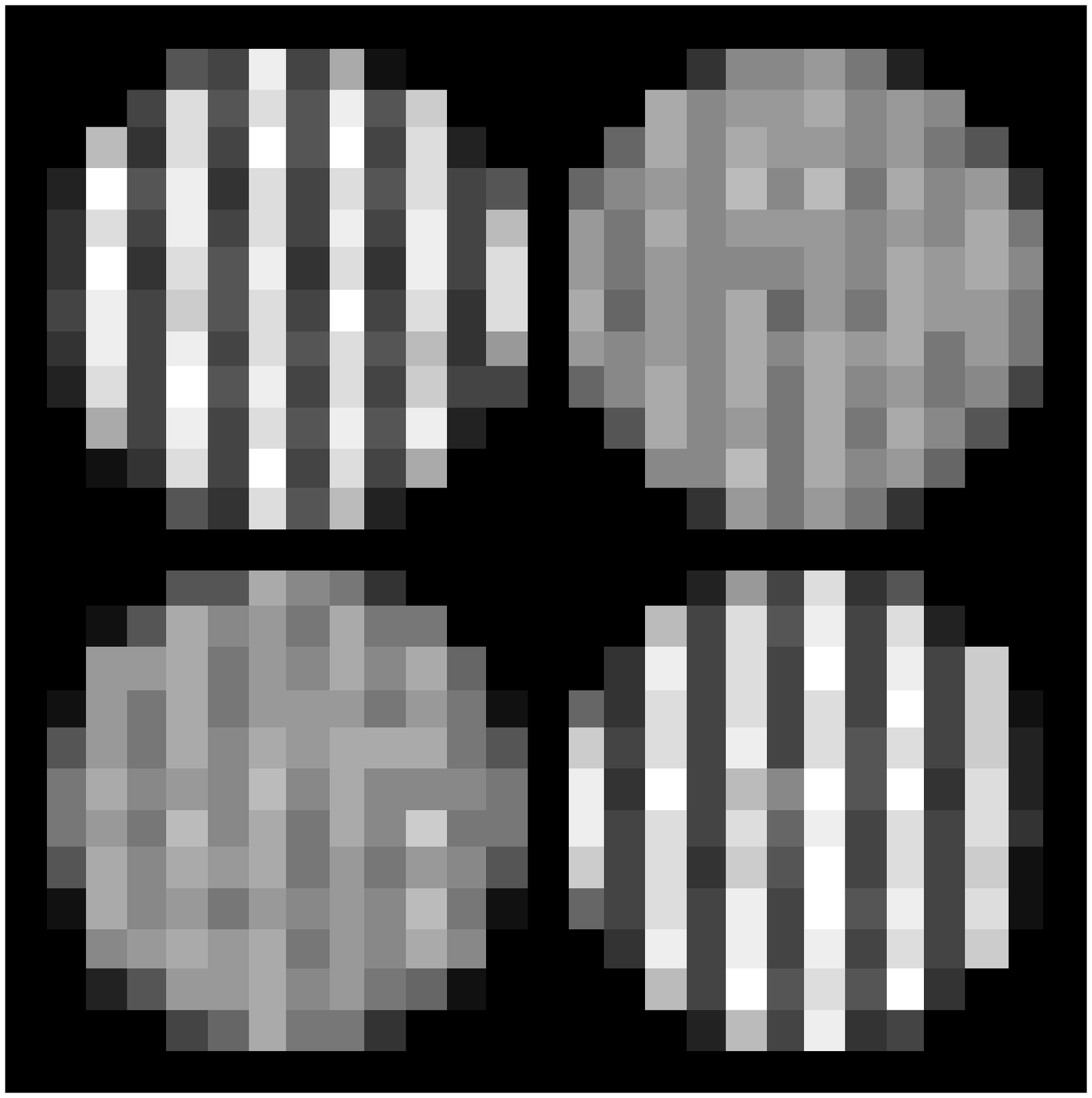}
\includegraphics[height=2.0in,width=2.0in]{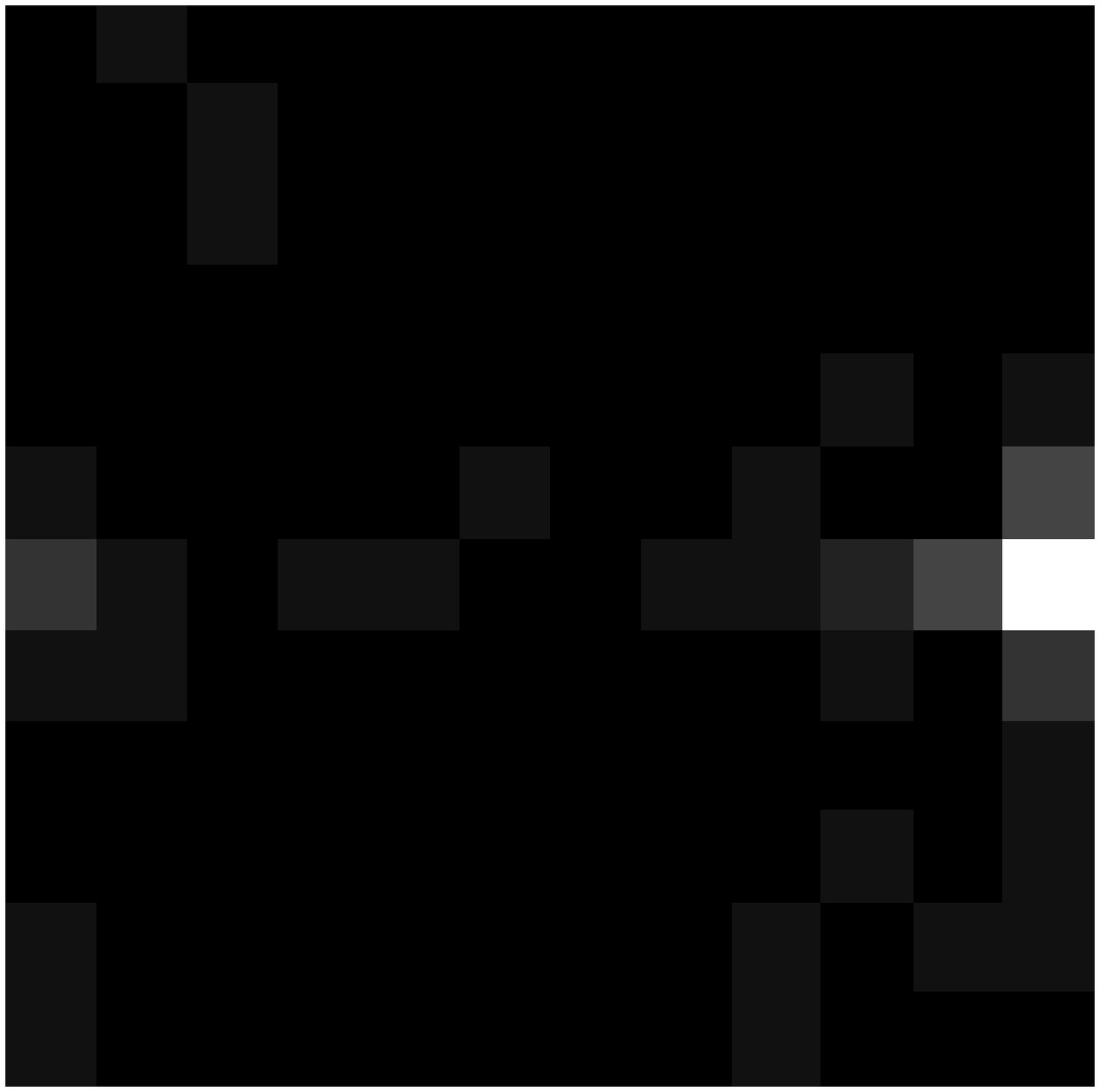}
\caption{(a) Fringes obtained with four pairs of zone plates and CZT for a source along 
X-axis with an offset of $204.5$ arcsec and (b) reconstructed image from the fringes. }
\end{figure}

In order to study the ability of the CZT detectors to resolve close-by sources, we now simulate
with two sources. The zone plate separation is $30$ cm. In this case, the angular resolution 
should be about $69$ arcsec. Two sources are placed at 
$\phi = 30$ and $99$ arcsecs respectively along X-axis ($\theta = 0$ degree). In Figs. 20(a-b) we
present the corresponding fringe patterns and the reconstruction of the sources.

\begin{figure}[h]
\centering
\includegraphics[height=2.0in,width=2.0in]{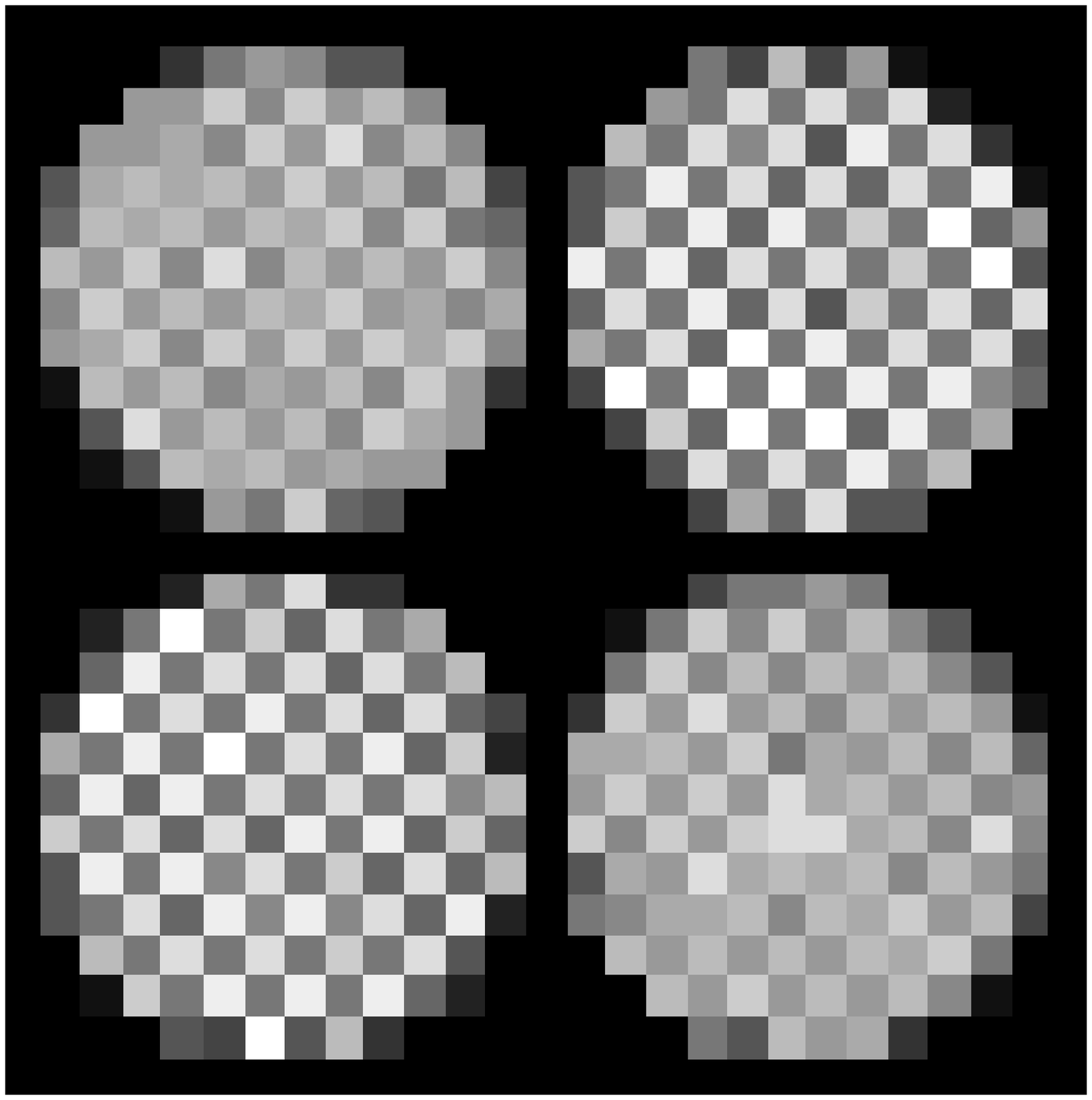}
\includegraphics[height=2.0in,width=2.0in]{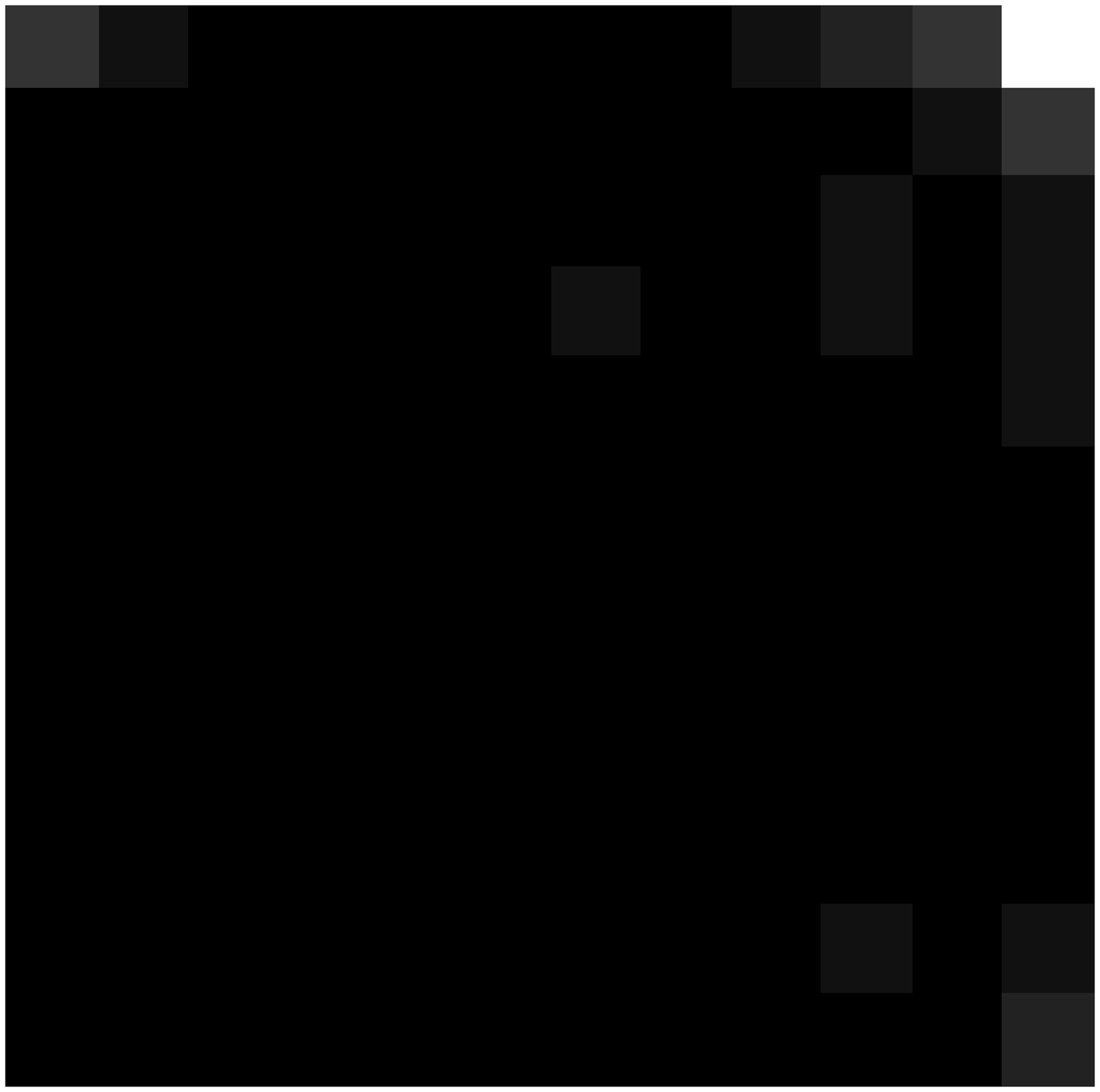}
\caption{(a) Fringes obtained with four pairs of zone plates and CZT for a source along
diagonal  with an offset  of $289.5$ arcsec and (b) reconstructed image from the fringes. }
\end{figure}

\begin{figure}[h]
\centering
\includegraphics[height=2.0in,width=2.0in]{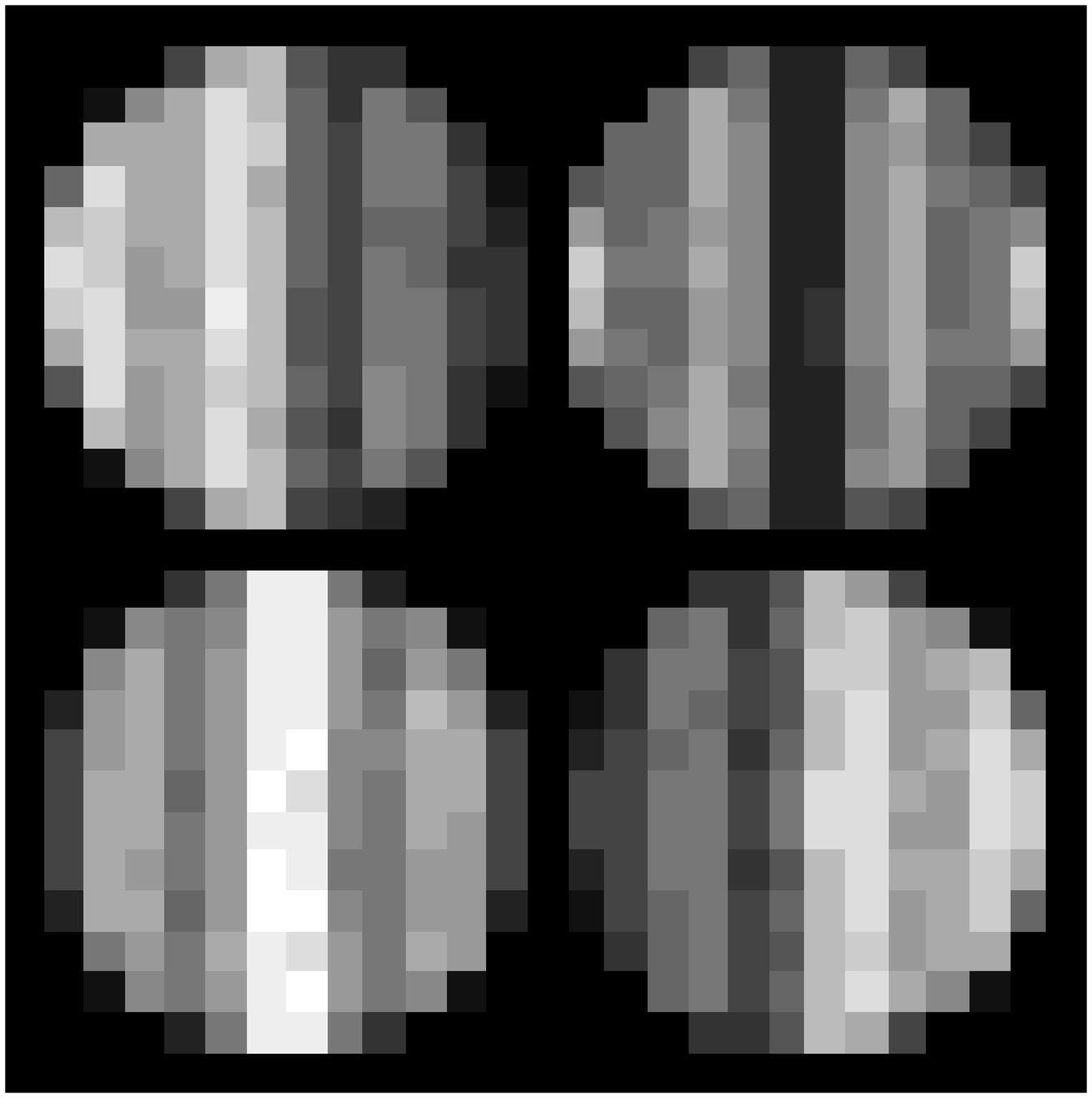}\hspace{0.5 cm}
\includegraphics[height=2.0in,width=2.0in]{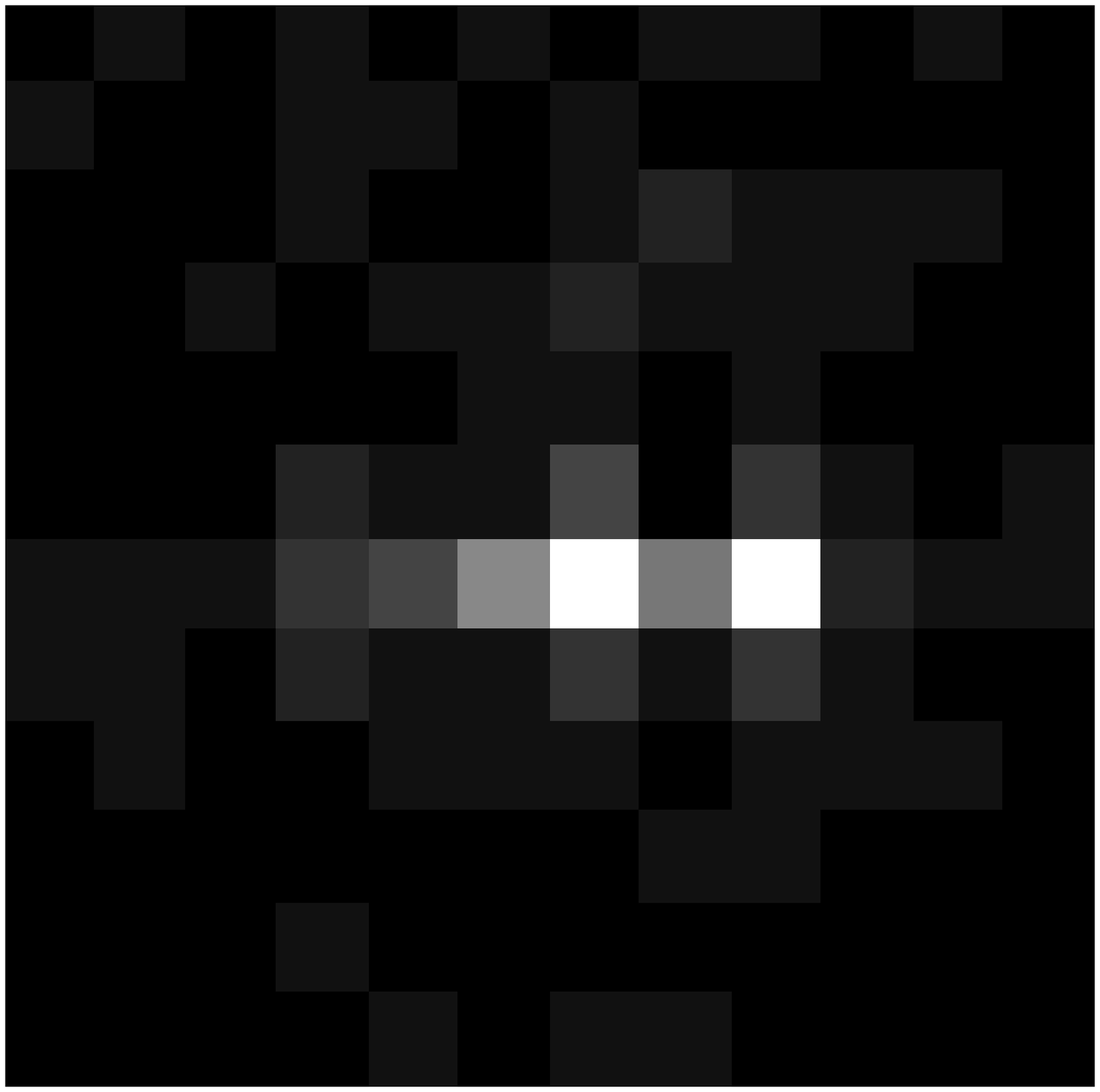}
\caption{(a) Fringes obtained with four pairs of zone plates and CZT detectors 
when two sources separated by $69$ arcsec are kept along the horizontal axis and 
(b) reconstructed images of the sources.}
\end{figure}

\section{Concluding remarks}

In this Paper, we presented results of various numerical simulations obtained by
varying the source, the zone plate telescope parameters and the detector parameters. 
We considered both two-pair and four-pair configurations which enabled us to 
remove the pseudo-source and both the pseudo-source and DC-offset
respectively. We pointed out that the pseudo-source cannot be totally removed if the 
source itself is at a finite distance. This is due to the fact that at a finite distance,
different pair is hit by photons from the source at various angles and the contributions
to the pseudo-source do not cancel out. We showed that the resolution of the instrument
is worsened when the source is placed at a finite distance. However, we showed that if the telescope
is modified in a way that both the plates subtend equal angles at the source, the high resolution
is recovered. Of course, a practical difficulty of such a method is that the second zone plate 
has to be modified dynamically as the source distance is varied, though, for instance for medical
purposes, one could keep the source at a fixed distance always and thus a fixed sized ZP2
would suffice. Alternatively, one could convolve the distorted fringes obtained due to sources
at a finite distance by theoretical Moir\'{e} pattern obtained for the source at infinite distance
before deconvolving the pixel information. This is beyond the scope of the present paper and will
be dealt with elsewhere.

We studied the cases with both CMOS and CZT detectors
and showed that the setup with CZT detectors will have very limited field of view in order
to even reconstruct a single source. This is because of the large pixel size. With a CMOS 
detector we simulated how a zone plate telescope would view the center of the Galaxy. When 
there are multiple sources of varying intensity, we showed that one could first obtain a
general image and then improve upon it by subtracting the fringes from the strongest ones.

The zone plate telescopes have very high potential for future space astronomy, especially since
it will have higher resolutions over a large range of energy. They may also be used for 
medical science since the sources at finite distances can also be resolved well if modified ZPTs are used.
Recently ZPTs have been used 
in RT-2/CZT payload aboard Russian satellite CORONAS-PHOTON for the first time. The 
satellite has been recently launched and the results are expected in near future,
especially when the sun becomes active.  The results of the zone plate imagers will 
be discussed elsewhere (Nandi et al. 2009).

\begin{acknowledgements}
We thank Dr. U. Desai for many helpful suggestions. SP and DD thank CSIR for supporting 
their research work.
\end{acknowledgements}



\end{document}